\documentclass[preprint,3p,authoryear,a4paper]{elsarticle}
\usepackage{graphicx}
\usepackage{epstopdf}
\usepackage{slashbox}
\usepackage{color}
\usepackage{dsfont}
\usepackage{amsmath}
\usepackage{amssymb}
\usepackage{amsfonts}
\usepackage{amsbsy}
\usepackage{caption}
\usepackage{subcaption}
\usepackage{amsthm}
\usepackage{array}
\usepackage{booktabs} 
\usepackage{verbatim}
\usepackage[english]{babel}
\usepackage{natbib}
\usepackage{ulem}
\usepackage{url}
\usepackage{multirow}
\usepackage{subcaption}
\usepackage{float}
\usepackage{graphics}

\renewcommand\appendix{\par
	\setcounter{section}{0}
	\setcounter{subsection}{0}
	\setcounter{table}{0}
	\setcounter{figure}{0}
	\gdef\thetable{\Alph{table}}
	\gdef\thefigure{\Alph{figure}}
	\gdef\thesection{\Alph{section}}
	\setcounter{section}{0}}

\newtheorem{remark}{Remark}[section]
\numberwithin{equation}{section}

\newcommand{\rev}[1]{\textcolor{black}{#1}}

	\newcounter{arclist}

		\newcounter{arcenum}

\begin{document}

				\normalem
				
				\begin{frontmatter}
					
					
\title{Detection and treatment of outliers for multivariate robust loss reserving}
					
\author[UMelb]{Benjamin Avanzi}
\ead{b.avanzi@unimelb.edu.au}
					
					\author[UNSW]{Mark Lavender\corref{cor}}
					\ead{mark.lavender@live.com.au}
					
					\author[UNSW]{Greg Taylor}
					\ead{gregory.taylor@unsw.edu.au}
					
					\author[UNSW]{Bernard Wong}
					\ead{bernard.wong@unsw.edu.au}
					
					\cortext[cor]{Corresponding author. }
					
\address[UMelb]{Centre for Actuarial Studies, Department of Economics, University of Melbourne VIC 3010, Australia}
\address[UNSW]{School of Risk and Actuarial Studies, UNSW Australia Business School, UNSW Sydney NSW 2052, Australia}

\begin{abstract}
Traditional techniques for calculating outstanding claim liabilities such as the chain ladder are notoriously at risk of being distorted by outliers in past claims data. Unfortunately, the literature in robust methods of reserving is scant, with notable exceptions such as \citet{VeDe11} and \citet{VeVa11}. In this paper, we put forward two alternative robust bivariate chain-ladder techniques to extend the approach of \citet{VeVa11}. The first technique is based on Adjusted Outlyingness \citep{HuVe08} and explicitly incorporates skewness into the analysis whilst providing a unique measure of outlyingness for each observation. The second technique is based on \textit{bagdistance} \citep{HuRoSe16} which is derived from the bagplot however is able to provide a unique measure of outlyingness and a means to adjust outlying observations based on this measure.

Furthermore, we extend our robust bivariate chain-ladder approach to an N-dimensional framework. The implementation of the methods, especially beyond bivariate, is not trivial. This is illustrated on a trivariate data set from Australian general insurers, and results under the different outlier detection and treatment mechanisms are compared.

\end{abstract}

\end{frontmatter}
{\centering \large}

\setcounter{page}{1}

\section{Introduction}\label{intro}
\subsection{Motivation}
At any moment, the number, timing and severity of future claims payments for a general insurer is shrouded in uncertainty. Reserves are set up to ensure, with a given degree of confidence, that necessary claims payments are met as they arise. The reserving problem is one of solvency, however it is also one of capital efficiency; if an insurer holds reserves over and above what is needed, they forfeit the opportunity to utilise this capital elsewhere, and insurance becomes more expensive than necessary. It is critical that the results from models and techniques applied to the loss reserving problem are as accurate as possible when tasked to a range of different data sets. 

Some of these data sets may include abnormal observations; outliers or deviations from model assumptions. Full inclusion of these data points in an analysis may prove detrimental to the accuracy of reserve estimates and the resulting inference. This is an issue that needs to be addressed if these models and techniques are going to reflect reality and be used to inform decisions. 
\emph{Robustness} refers to the ability of a model or estimation procedure to not be overtly influenced by outliers in the data set under investigation and/or deviations from the underlying assumptions of the model. Understanding how outliers may impact the results of a model and having statistically sound procedures to detect and treat abnormal observations will improve the robustness of reserving techniques and ultimately lead to more informed and reliable decisions. 

While some authors have explored the issue of robustness in reserving, the body of literature in this area is relatively scant. Of particular importance for this paper is the robust GLM chain-ladder of \citet*{VeDe11} and the robust bivariate chain-ladder of \citet*{VeVa11}, however there has been notable work that moves away from the chain-ladder technique. For robust non-chain ladder methods, please also see for example \cite{ChCh03,ChChMa08,PiGrBa15}.

\citet*{VeVaDh09} provide a two-stage deterministic robust chain-ladder technique which fundamentally relies on the analysis of residuals given after fitting an over-dispersed Poisson (ODP) GLM to the cumulative and then incremental claims data as described in \citet*{EnVe99}. A boxplot is employed on the Pearson residuals after fitting the ODP GLM at each stage to detect outlying observations.  Under their robust technique, development factors are calculated as medians which are known to be much more robust than means. Unfortunately, in its original formulation, results were poor as reserves were still being heavily influenced by outliers.

The approach of \citet*{VeVaDh09} was refined by \citet*{VeDe11}. In particular, it was found that the standard threshold value of 1.345 to identify outliers was often too low for triangular loss data. To combat this an additional stage was added to the methodology whereby the threshold point was taken to be the 75\%-quantile of the residuals after an initial fit using the threshold value of 1.345 
which led to better results in terms of robustness. \citet*{VeDe11} also showed that the influence function for reserves with respect to incremental claims is unbounded when assuming a Poisson GLM specification. This provides a formal basis for the non-robustness of the chain-ladder technique, and related techniques. A comprehensive study of impact functions of central estimates, mean squared errors, and quantiles of reserves can be found in \citet*{AvLaTaWo23}.

\subsection{Statement of contributions}
The refined robust GLM chain-ladder technique is an integral component of the robust bivariate chain-ladder as developed in \citet*{VeVa11}. In particular, the residuals given after fitting the robust GLM chain-ladder are used to generate a bagplot \citep*{RoRuTu99} which may be considered as a bivariate boxplot. The second approach employed to detect and treat outliers is the minimum covariance determinant (MCD) \citep*{Rou84} Mahalanobis distance, also applied to the residuals. 

Each of these two techniques have shortcomings that this paper aims to address. In particular, they fail to take skewness of the data sufficiently into account, which may lead to the misclassification of data points between regular and outliers. 

We introduce two alternative methodologies that offer a consistent and structured approach to the detection, measurement and treatment of outlying observations with statistical backing. Our methods are based on Adjusted Outlyingness \citep*{HuVe08} and \textit{bagdistance} \citep*{HuRoSe16}. Adjusted Outlyingness provides a unique measure of outlyingness for each observation and explicitly incorporates a robust measure of skewness into the detection process. \textit{Bagdistance} is derived from the bagplot and provides a measure of outlyingness for each observation. Through calculation of the \textit{bagdistance} a greater variety of alternative treatments for outliers becomes available then when simply using the bagplot. These methodologies are applied and compared on real data. Improvements over the techniques introduced in \citet{VeVa11} are shown to be material and significant.

Additionally, we extend the methodologies beyond the bivariate case. Multivariate reserving techniques primarily benefit from the consideration of the dependence structure between loss triangles. Such consideration should provide a more accurate estimation of reserves, however the impact of outliers can still be significant in this setting. Furthermore, information about other lines of business may inform the identification and adjustment of outliers. Hence appropriate detection techniques, and adjustment procedures help ensure that the benefits of multivariate reserving analysis are not lost due to a lack of robustness. This is also a focus of this paper, which develops a multivariate robust reserving technique. We extend the bivariate chain-ladder to an N-dimensional framework and illustrate the implementation of all four outlier detection and treatment techniques on a trivariate example. 

Implementation of the methodology is not trivial, and R codes are freely available on GitHub for any interested user; see Section \ref{S_Rcode}.


\section{Outlier Detection Techniques}\label{bivarout}

In Section \ref{S_prel} we start by reviewing some basic concepts related to robust statistics, which are required to introduce four detection techniques in Section \ref{S_ODT}.

\subsection{Preliminary: General Robustness Concepts} \label{S_prel}

\subsubsection{Influence Functions and Breakdown Points}

When considering robust methodologies, two common concepts are influence functions and breakdown points, which we define here for later reference:

\textbf{Influence functions}  consider an estimation technique and calculate the potential `influence' a single data point could have on that estimator. For robustness it is desirable to have a bounded influence function such that no single data point can have an unlimited impact on estimators. An example of an estimator with an unbounded influence function is the mean whereas the median has a bounded influence function. 

\textbf{Breakdown points} describe the proportion of the data set required to be contaminated (or outlying, however this may be defined) before an estimator provides inaccurate results. For example, when calculating the mean only one data point has to be contaminated to render the technique invalid, however, when calculating the median, 50\% of data points need to be contaminated to invalidate the technique.

\subsubsection{Halfspace Depth and Tukey Median} \label{hddescription}

Both the bagplot \citep*{RoRuTu99} and the bagdistance (\emph{bd}) \citep*{HuRoSe16} are based on the concept of halfspace depth \citep*{Tuk75}.

\begin{figure}[htb]
	\centering 
		\centering
		\includegraphics[width=0.6\linewidth]{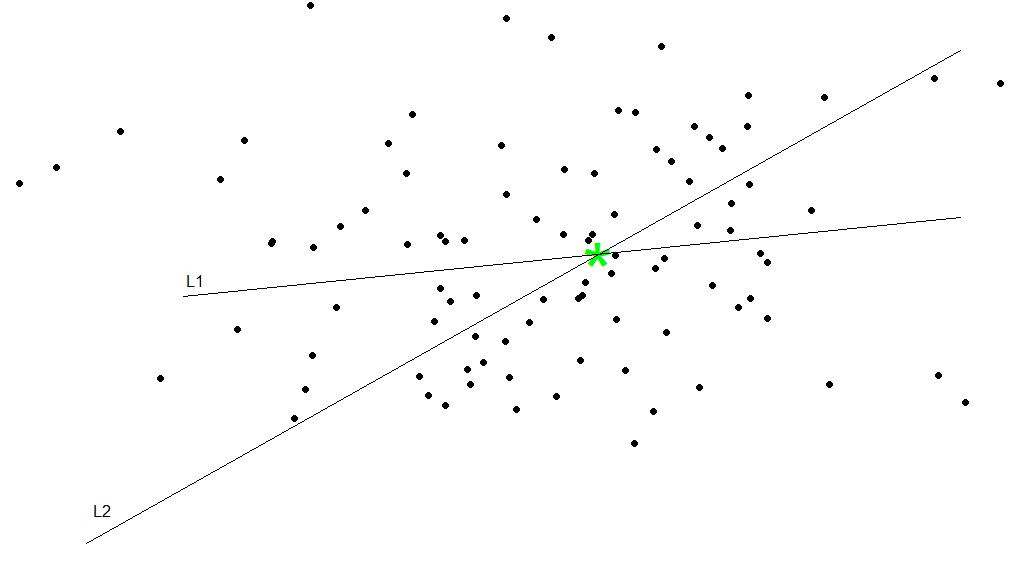}
		\caption{Halfspace Depth Illustration in 2 Dimensions}
	\label{halfspace}
\end{figure}

The \textbf{halfspace depth} \citep*{Tuk75} of a point is defined as the minimum number of points (from the data sample) in a closed halfplane through the point of interest. We refer to Figure \ref{halfspace} which illustrates the concept of halfspace depth in the bivariate case (note that this is scalable to higher dimensions).
  In Figure \ref{halfspace}, to calculate the halfspace depth of the point marked with a green asterisk, we would consider numerous lines through this point (such as the lines denoted as $L_1$ and $L_2$ in the Figure) and look for the minimum number of points on either side of each of these lines. \citet*{Rom01} showed that halfspace depth has a \emph{bounded influence function}. 
  This means that the impact an outlier may have on the halfspace depth of a given observation is limited and highlights the robustness of this statistic.
  
  The \textbf{depth median (Tukey median)} \textbf{{T}$^*$}, is the point with the greatest halfspace depth. It represents a central point of the data and is marked by the red asterisks in Figure \ref{bpill}. If the point is not unique the centre of gravity of the deepest region is used.
  
\subsubsection{Minimum Covariance Determinant (MCD) Estimation Procedure} \label{MCDEstimation}
A popular technique to estimate location and dispersion parameters in a robust fashion is the \textbf{minimum covariance determinant (MCD)} \citep*{Rou84} procedure, whereby one finds the 
\begin{equation}
    \left\lfloor \frac{n+p+1}{2}\right\rfloor\leq h\leq n
\end{equation}
observations whose classical estimator of the covariance matrix has the smallest determinant where $n$ represents the number of observations and $p$ represents the dimension of the data. \rev{The variable $h$ represents the number of observations that are ultimately used, and the procedure puts conditions around the minimum number of observations to avoid selection of a small set that doesn’t truly represent the data.} The location vector is then the arithmetic mean of these points and the scale matrix is taken as a multiple of this covariance matrix. Heuristically, this procedure identifies a `core' set of data points which are mostly related to one another. 

\subsubsection{Swamping and Masking}

Inappropriate outlier detection techniques may lead to swamping or masking. \textbf{Swamping} occurs when one classifies too many observations as outliers i.e. the impact outliers have on the technique being applied results in regular observations incorecctly appearing to be outliers. \textbf{Masking}, on the other hand, refers to the failure to classify outliers as such i.e. outliers have impacted the technique being applied in a way that 'masks' them from being detected as outliers.

\subsection{Outlier Detection Techniques} \label{S_ODT}

A robust bivariate chain-ladder technique hinges on the detection and adjustment of bivariate outliers. \citet*{VeVa11} put forward two techniques to be used for this task: the MCD \citep*{Rou84} \textbf{Mahalanobis Distance}, and the \textbf{bagplot} \citep*{RoRuTu99}. In this section we outline these approaches and two different techniques; the \textbf{bagdistance} and \textbf{Adjusted Outlyingness}---that can be applied to the multivariate reserving problem and which address some of the shortcomings of the former two.

\subsubsection{MCD Mahalanobis Distance}

A standard method used to detect outliers in multivariate analysis is to calculate some measure of distance of each data point from the centre of the data. A popular measure is the \textbf{Mahalanobis Distance (MD)}, given by, 
\begin{equation}
    \text{MD}(\textbf{x}_i)=\sqrt{(\textbf{x}_i-\widehat{\pmb{\mu}})'\widehat{\pmb{\Sigma}}^{(-1)}(\textbf{x}_i-\widehat{\pmb{\mu}})},
\end{equation}
where $\widehat{\pmb{\mu}}$ represents the sample location vector and $\widehat{\pmb{\Sigma}}$ represents the sample scale matrix. If these are not estimated in a robust manner then outliers may fail to be detected due to masking and swamping effects. Essentially this means that outlying observations will influence the estimate of the central point and dispersion of the data leading to outliers themselves having small MD (masking) and/or non-outlying observations having high MD (swamping).
To combat this, the MCD estimates of location and dispersion are used, and outliers are then flagged as observations that have a MD greater than a certain threshold. Note that $\text{MD}^2\sim \chi_p^2$ if the underlying data is normal.

\subsubsection{Bagplot}\label{S_bagplot}
\begin{figure}[htb]
	\centering 
		\centering			 	
		\includegraphics
		[width=0.7\textwidth]{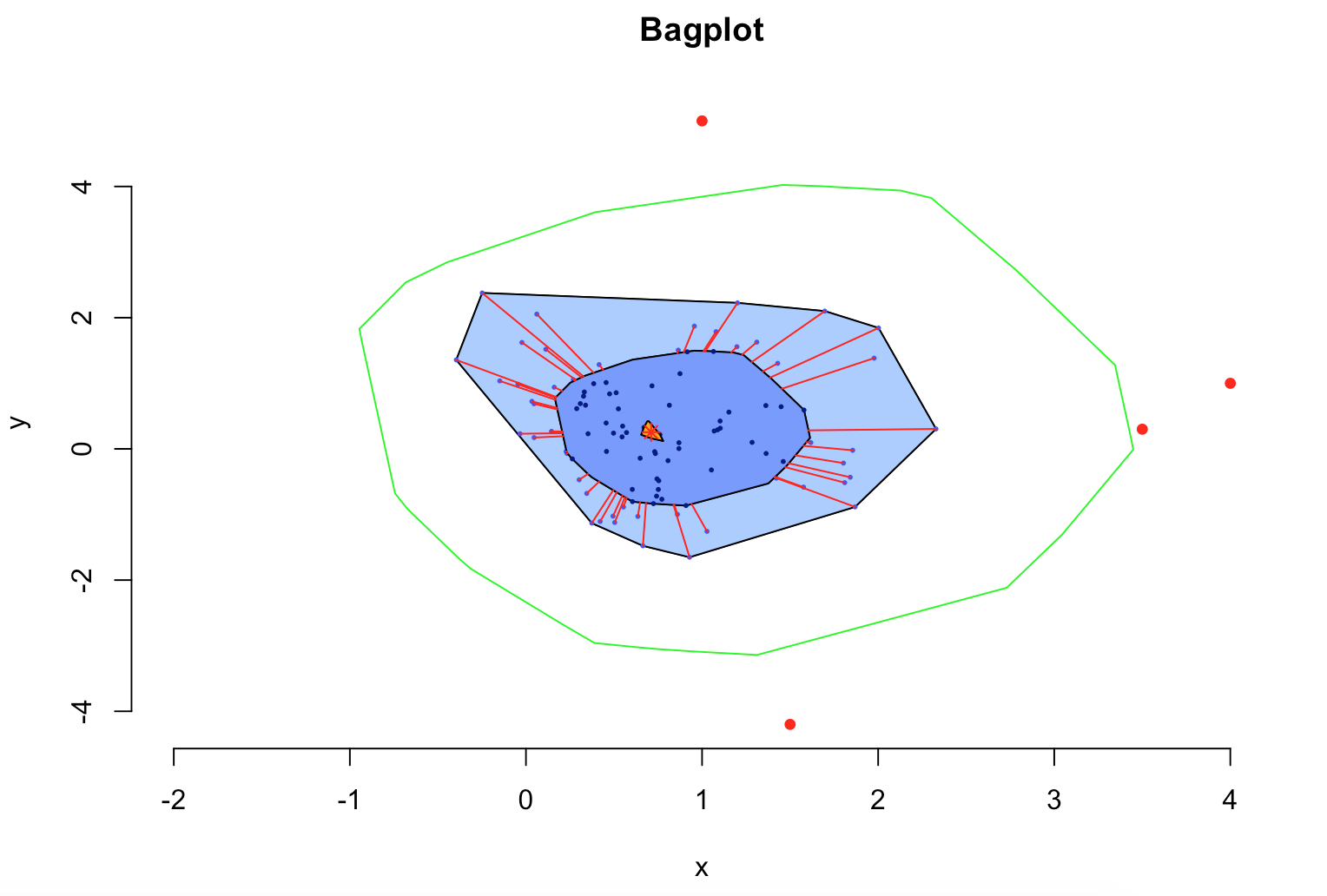}
		\caption{Bagplot with Fence Drawn} 
	\label{bpill}
\end{figure}

We refer to Figure \ref{bpill} to outline the components of a \textbf{bagplot}. 

The \textbf{bag} \textbf{(B)}, is shown in Figure \ref{bpill} by the darker inner area surrounding \textbf{T}* and is constructed as follows. First, denote $D_k$ as the region of all data points that have halfspace depth greater than $k$ and $\#D_k$ as the number of data points contained in this region. The bag is given by the linear interpolation with respect to \textbf{T}$^*$ of the two regions that satisfy $\#D_k\leq \lfloor \frac{n}{2}\rfloor<\#D_{k-1}$. Here linear interpolation refers to the method of constructing new data points that are in the range of both of $\#D_k$ and $\#D_{k-1}$. 

The \textbf{fence} is then given by multiplying the bag by some factor relative to \textbf{T}$^*$. Typically this factor is three and \citet*{RoRuTu99} point out that this value was chosen based on simulations. However, in their recent R package, mrfDepth, \cite*{mrfDepth} have instead opted to use the square root of the  99$^{th}$ percentile of a chi-squared distribution where the degrees of freedom is equal to the number of dimensions being considered. This new fence factor was chosen after derivation of the distribution of the \textit{bagdistance} \citep*{HuRoSe15}; see also the next section. In the two dimensional case it is very close to three. We will use this newly developed fence factor throughout the paper. 

When displaying a bagplot, the fence is generally not drawn \citep*{RoRuTu99}. 
Figure \ref{bpill}, however, shows the same bagplot with the fence drawn in green and we can see that the four outlying points marked in red are outside this fence. Note that the choice of this fence factor will directly influence both the number of outliers detected and how they are adjusted. Data points outside the fence are considered outliers and are adjusted to facilitate the application of loss reserving techniques. The adjustment may be done in a purely graphical manner i.e. bringing outliers back to the fence or loop or a weighting function based on \textit{bagdistance} may be employed.
 
 The perimeter of the lighter blue area is know as the \textbf{loop} and is given by the convex hull of all non-outlying points. This is analogous to the whiskers in a univariate boxplot \citep*{Tuk77}.
 
\subsubsection{Bagdistance}\label{bagd}
Now we present the \textbf{bagdistance (\textit{bd})} \citep*{HuRoSe16}. This statistic provides a measure of outlyingness for each observation and hence does not rely solely on graphical representations. It can also handle skewness often found in loss data. \rev{However, }similarly to the bagplot, the \textit{bd} utilises the bag to capture the shape of the data and subsequently detect outliers. As the bag is formulated based on circa 50\% of the data points, there is potential that it does not fully encapsulate the skewness in the set. When this is the case, the ensuant outlier detection results may be flawed. To calculate the \textit{bd}, firstly define \textbf{c}$_x$ as the intersections of the boundary of the bag (B) and the ray from the Tukey median \textbf{T}$^*$ through the points $\textbf{x}$ (red lines in Figure \ref{bpill}). The \textit{bd} is defined as follows 
\begin{equation}
bd(\textbf{x};P_n)=
\begin{cases}
{0}\ \text{if}\ \textbf{x}=\textbf{T}^*;\\
\frac{||\textbf{x}-\textbf{T}^*||}{||\textbf{c}_x-\textbf{T}^*||}\ \text{elsewhere},
\end{cases}
\end{equation}
where $P_n$ represents the distribution of the data set and $||.||$ is the Euclidean norm such that $||\textbf{x}||=\sqrt{x_1^2+...+x_n^2}$.
The denominator scales the distance of the data point (\textbf{x}) to \textbf{T}$^*$, relative to the dispersion of the bag (in that direction of the projected ray). A cut-off point is then set, such that data points with a \textit{bd} beyond this threshold are considered outliers and are adjusted back to an appropriate point on the ray emanating from \textbf{T}$^*$ passing through \textbf{x}. Note that if the cutoff value is chosen to be the same as the fence factor under the bagplot technique, then the same outliers will be detected.

\subsubsection{Adjusted Outlyingness}\label{adjoutdescription}

\textbf{Adjusted Outlyingness} \citep*{HuVe08} is based on an adjustment of the Stahel- Donoho estimate of outlyingness \citep*{Don82} to explicitly incorporate skewness. Furthermore, it is based on robust estimates of location, scale and skewness such that it achieves a theoretical breakdown point of 25\%. Further highlighting its robustness, the adjusted outlyingness technique has a bounded influence function \citep*{HuVe08}. The technique is applied by firstly considering a $p$-dimensional sample $\textbf{X}_n=(\textbf{x}_1, ... ,\textbf{x}_n)'$ where $\textbf{x}_i=(x_{i1},...,x_{ip})'$ and $\textbf{a}\in \mathbb{R}^p$. The measure of adjusted outlyingness (AO) for $\textbf{x}_i$ is given by 
\begin{equation}
AO_i=	\sup_{\textbf{a}\in \mathbb{R}^{p}}AO^{(1)}(\textbf{a}'\textbf{x}_i,\textbf{X}_n\textbf{a}),
\end{equation}
where \begin{equation}
AO^{(1)}(x_i,X_n)=\begin{cases}
\frac{x_i-\text{med}(X_n)}{w_2-\text{med}(X_n)},& \text{if } x_i>\text{med}(X_n), \\
\\ \frac{\text{med}(X_n)-x_i}{\text{med}(X_n)-w_1}, &\text{if } x_i<\text{med}(X_n),
\end{cases}
\end{equation}
 where $w_1$ and $w_2$ are the lower and upper whiskers of the skew-adjusted boxplot \citep*{HuVa08} applied to the data set $X_n$, and where $\text{med}(X_n)$ denotes the median of that data set. The skew-adjusted boxplot for all the AO-values is then constructed and those that are beyond the cut-off value \begin{equation}\label{AOcut}
\text{cut-off}=Q_3+1.5e^{3MC}IQR \quad
\end{equation}
are declared as outliers, where $Q_3$ represents the third quartile of the data, MC represents the medcouple \citep*[a robust measure of skewness; see][]{BrHuSt04}, and where IQR is the interquartile range (i.e. $Q_3-Q_1$, where $Q_1$ represents the first quartile of the data). We are only concerned with the upper cut-off value as we are performing the skew-adjusted boxplot technique on the measures of outlyingness and hence small results in this context are not of interest.
An alternative AO cut-off value of $\sqrt{\chi^2_{\{99,p\}}}\cdot\text{median}(AO)$ is given in mrfDepth \citep{mrfDepth}, where $p$ represents the dimension of the data, $\chi^2_{\{99,p\}}$ is the 99$^{th}$ percentile of the chi-squared distribution with $p$ degrees of freedom and AO represents the set of all AO values for the data set. This cut-off value aligns closely with the fence factor used for the bagplot approach. In this paper we consider both cut-off values.

Under the AO methodology, not all univariate vectors \textbf{a} can be considered, however \citet*{HuVe08} note that taking $m=250p$ directions provides a good balance between `efficiency' and computation time. From here, if $p=2$, to visualise the bivariate data a version of the bagplot based on the AO values (rather than halfspace depth) may be constructed \citep*{HuVe08}. 


When constructing an AO based bagplot, the bag is given by the convex hull of the 50\% of data points with the smallest AO (note this is different from Figure \ref{bpill}, which was based on halfspace depth). There are three possible approaches, which differ in the mechanism used to detect and hence treat outliers:
\begin{enumerate}
    \item A fence is drawn by multiplying the AO based bag by 3 (or some other factor) relative to the point with the lowest AO. Outliers are then flagged as those observations outside the fence and the loop is the convex hull of all points within the fence.
    
    Since the fence is generated from the bag which only considers 50\% of the data points it may fail to fully capture the shape of the data and in particular the skewness in the set.
    \item Utilise the alternative cut-off value given in mrfDepth where no fence is drawn.
    
    Here, the cut-off does not incorporate a robust measure of skewness and instead relies on the median value of the AOs and a quantile of the chi-squared distribution.
    \item Outliers are flagged using the traditional cut-off value given by Equation \eqref{AOcut}. In this case, the loop will be generated by the convex hull of all points with AO less than this value and no fence will be generated.
    
    The traditional AO cut-off value  incorporates a robust measure of skewness known as the medcouple which considers the whole data set. Hence, it is more equipped to capture the total skewness, and the loop that is generated by this approach more fully captures the skewness in the data in comparison to the fence methodology.
\end{enumerate}
For the reasons explained above and unless stated otherwise, we will use the third approach.

\section{An extension of the robust bivariate chain-ladder}\label{robcl}
In this section we implement the Adjusted Outlyingness and \textit{bagdistance} (see Section \ref{bivarout}) outlier detection and treatment techniques in a bivariate reserving setting. 
We show that the four techniques often lead to different results (although the bagplot and bagdistance will be the same if you use the fence factor as the cut-off distance). The differences are not only in the number of outliers flagged but also which observations are detected and hence treated. In each case, the different techniques should be implemented and results compared. 

This section is motivated by some of the shortcomings of the current robust bivariate chain-ladder methodology. Firstly, the MCD Mahalanobis distance approach assumes elliptical symmetry of the multivariate data. If this assumption is not met, we may fail to detect outlying observations as well as falsely declare regular observations as outliers due to masking and swamping effects. The bagplot is better able to effectively visualise bivariate data; highlighting any correlation, skewness and tail behaviour. 

 We now present comparisons between these four outlier detection techniques when applied to real non-life insurance data.
\subsection{The Robust Bivariate Chain-Ladder} \label{S_VeVa11}

	The robust bivariate chain-ladder technique was put forward by \citet*{VeVa11} and the general steps involved are as follows: \begin{enumerate}
		\item Apply the robust Poisson GLM chain-ladder technique (Verdonck and Debruyne, 2011) to each triangle separately. Obtain residuals from each triangle given by 
		\begin{equation}
		r_{ij}=\frac{X_{ij}-\widehat{\mu}_{ij}}{V^{\frac{1}{2}}(\mu_{i,j})},
		\end{equation}
		where $X_{ij}$ represents the incremental claim for accident year $i$ and development year $j$.
		\item Store residuals from each triangle as bivariate observations in a $n\times 2$ matrix $\textbf{X}=(\textbf{x}_1,...,\textbf{x}_n)'$ where $\textbf{x}_i=(r_{kj1},r_{kj2})$. We have that $n=I(I+1)/2$.
		\item Apply either the bagplot or MCD Mahalanobis Distance (see Section \ref{bivarout}) to these bivariate residuals.  
		\item Adjust outliers. For the bagplot, outlying observations are brought back to the fence or loop. For the MCD Mahalanobis distance technique, observations are brought back to the tolerance ellipse representing the 95\% quantile of the $\chi_2^2$ distribution.
		\item Backtransform adjusted residuals to obtain robust incremental claims $X_{i,j}^{Rob}$. 
		\item Apply the multivariate time series chain-ladder \citep{MeWu08} to the robust (adjusted) observations. 
	\end{enumerate}
We now provide an example of the above framework on real bivariate data. We however include two alternative outlier detection and treatment techniques: the Adjusted Outlyingness and \textit{bagdistance}. 
\subsection{Bivariate Example using Belgian data}
	The data for this example is from Belgian non-life insurers and is given in Appendix \ref{bidata} \citep*[from][]{ShBaMe12}.

\subsubsection{Bagplot}
\begin{figure}[htb]
		\begin{center}
			\includegraphics[width=0.7\textwidth]{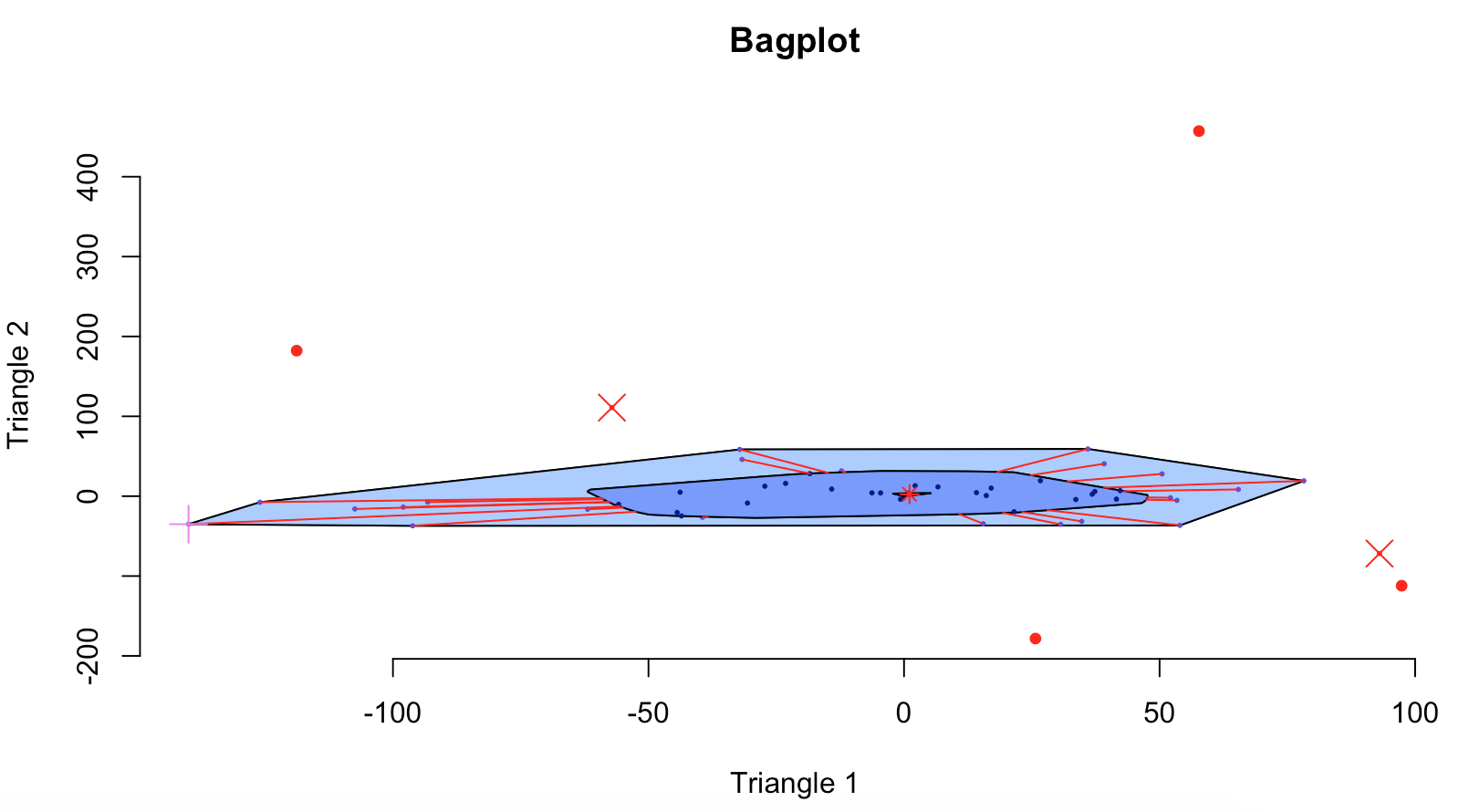} 
			\protect\caption{Bagplot} 
			\label{ShBaMe12HD=1}
		\end{center}
	\end{figure}

  We begin by illustrating the major  shortcoming of the bagplot approach in that it does not provide a measure of outlyingness that is unique for each observation. This is because the bagplot approach is based on halfspace depth whereby multiple different observations can have the same halfspace depth even if they are at different levels of outlyingness. Figure \ref{ShBaMe12HD=1} shows the bagplot for this data, where 6 outliers have been detected. Four of these observations have the lowest halfspace depth of 1 which may be expected for outliers. However, two of these outliers, $X_{6,5}$ and $X_{7,3}$, have a halfspace depth of 2 and 3 respectively. These points are marked with red crosses. 
 Additionally, one non-outlying observation, $X_{9,2}$ has a halfspace depth of 1 and is marked with a purple `+'.  Hence without the corresponding bagplot, little inference can be made about how outlying the outliers are and importantly about whether an observation is outlying or not. This will become an even bigger concern when extending this technique to higher dimensions where graphical representations become less available. Indeed, the bagplot is purely based on a form of ranking (halfspace depth) (see Section \ref{S_bagplot}). As a result, without having the graph available (such as for higher dimensions as considered in the next section) it is difficult to communicate how outlying an observation is.

\begin{figure}[htb]
	\centering 
	\begin{subfigure}{.5\textwidth}
		\centering
		\includegraphics[width=\textwidth]{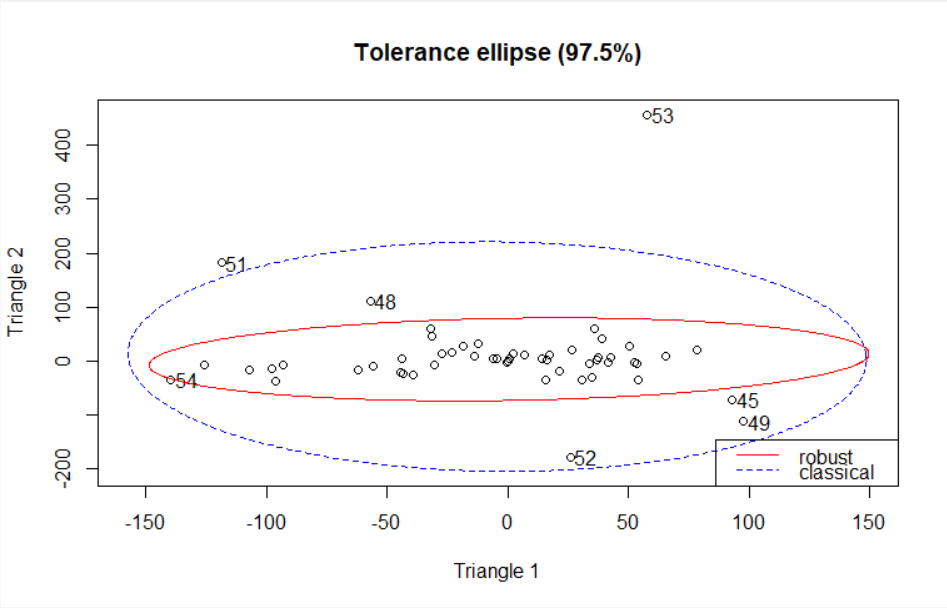}
		\caption{\rev{Tolerance Ellipses Before Adjusting Outliers}}
	\end{subfigure}%
	\begin{subfigure}{.5\textwidth}
		\centering			 	 		\includegraphics[width=\textwidth]{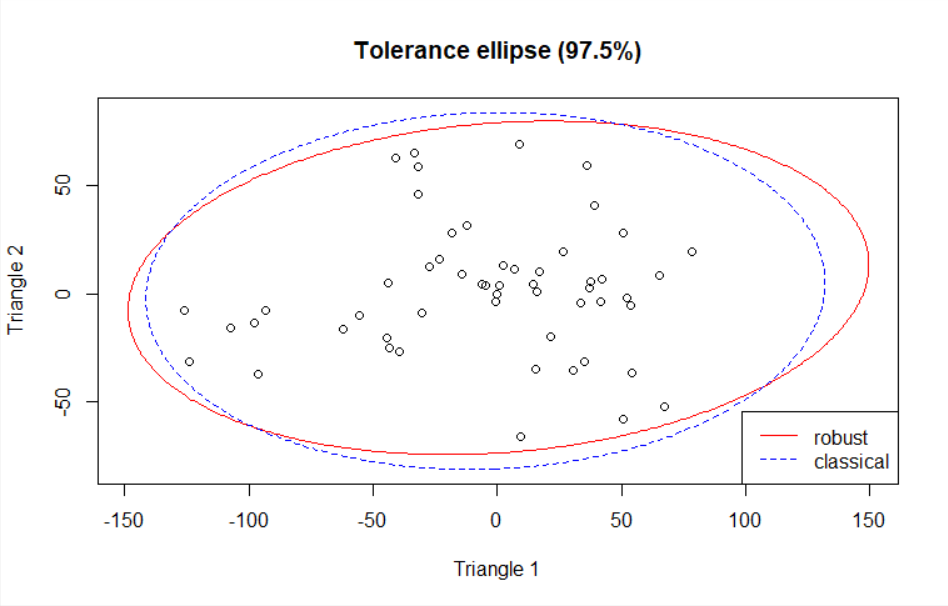}
		\caption{\rev{Tolerance Ellipses After Adjusting Outliers}}
	\end{subfigure}
	\caption{Tolerance Ellipses}
	\label{ShBaMe12toleranceellipse}
\end{figure} 

\subsubsection{MCD Mahalanobis Distance}\label{MCD}
When employing the MCD Mahalanobis \rev{Distance} technique a graphical representation is also available in the bivariate case. In particular, we can plot each data point as well as tolerance ellipses of which have a squared distance to the central estimate of the data equal to a quantile of the $\chi^2_2$ distribution. \rev{The classical tolerance ellipse is constructed when the Mahalanobis Distance is calculated using the classical estimators of the location vector and scale matrix rather than their robust counterparts from the Minimum Covariance Determinant procedure outlined in Section \ref{MCDEstimation}.} Outliers are adjusted by a technique known as \textbf{bivariate Winsorization} such that an outlying observation \textbf{x} is adjusted according to 
\begin{equation}
\text{min}\left(\sqrt{\frac{c}{MD^2(\textbf{x})}},1\right)\cdot\textbf{x},
\end{equation}
where $c$ is equal to the 95\% quantile of a $\chi^2_2$ distribution ($\chi^2_{0.95,2}$).  Under this methodology 7 observations are detected as outliers. These are the same observations as was flagged under the bagplot approach as well as $X_{9,2}$, the only other observation with a halfspace depth of 1. Figure \ref{ShBaMe12toleranceellipse} gives these plots before and after outliers have been adjusted.

\begin{figure}[htb]

			\centering
			\includegraphics[width=0.6\textwidth]{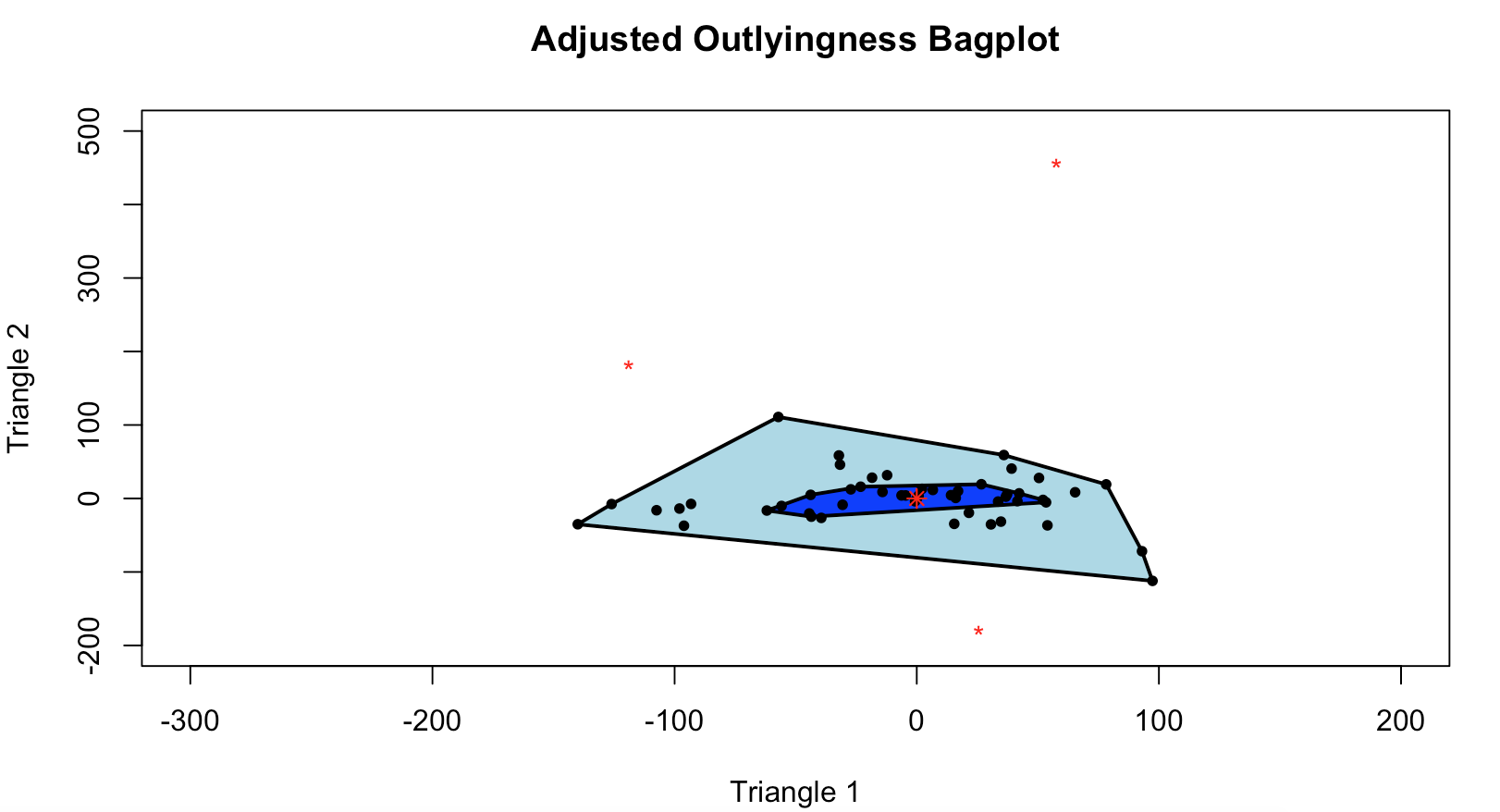}
			\caption{Adjusted-Outlyingness Bagplot Using Traditional Cut-off Value (Approach 3)}
			\label{ShBaMe12AObp}
\end{figure}

\begin{figure}
		\centering
		\begin{subfigure}{.45\textwidth}
			\centering				
		    \includegraphics[width=\textwidth]{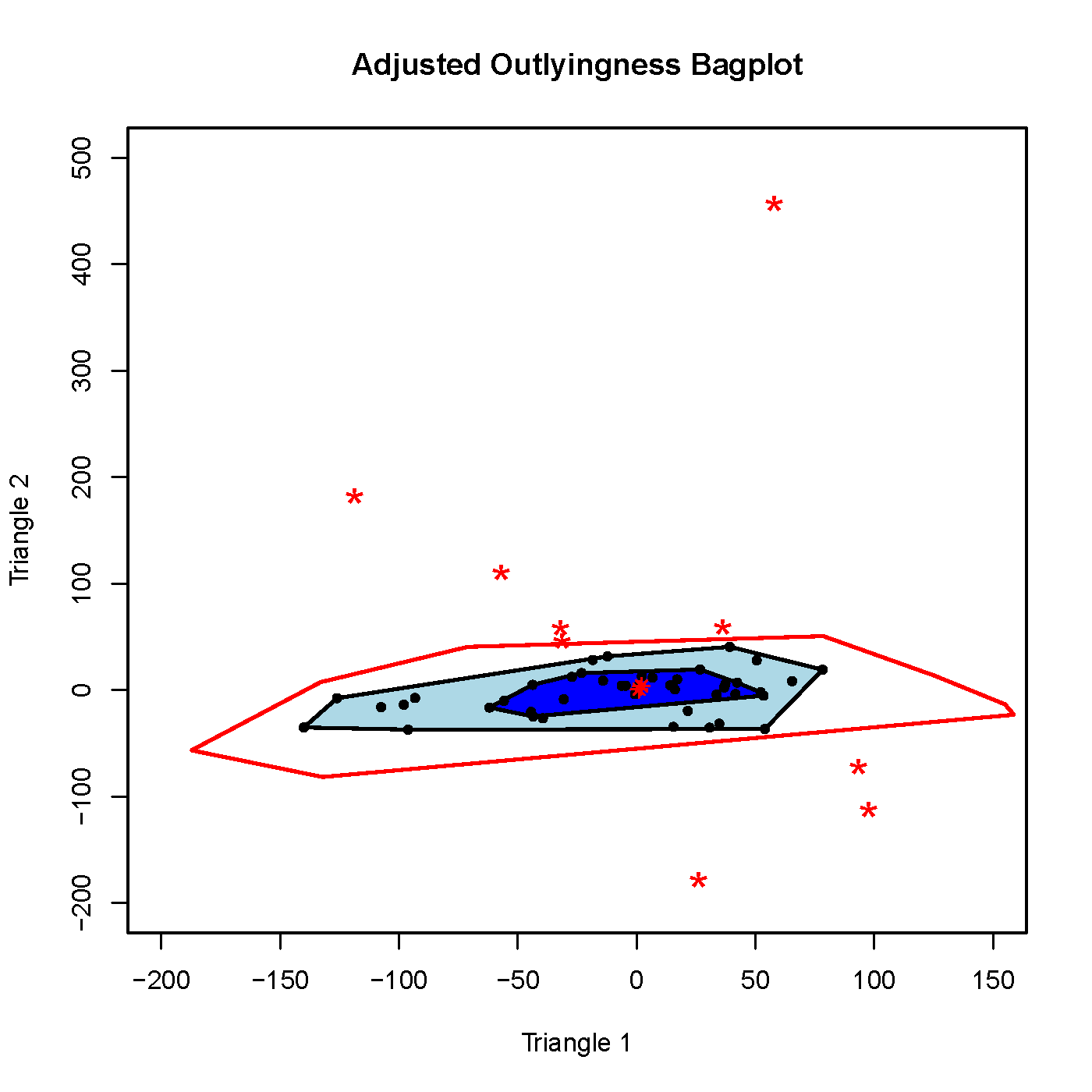}
			\caption{Adjusted-Outlyingness Bagplot With Fence (Approach 1)}
			\label{ShBaMe12AObpfence}
		\end{subfigure}
		\begin{subfigure}{.45\textwidth}
		\centering
 	\includegraphics[width=\linewidth]{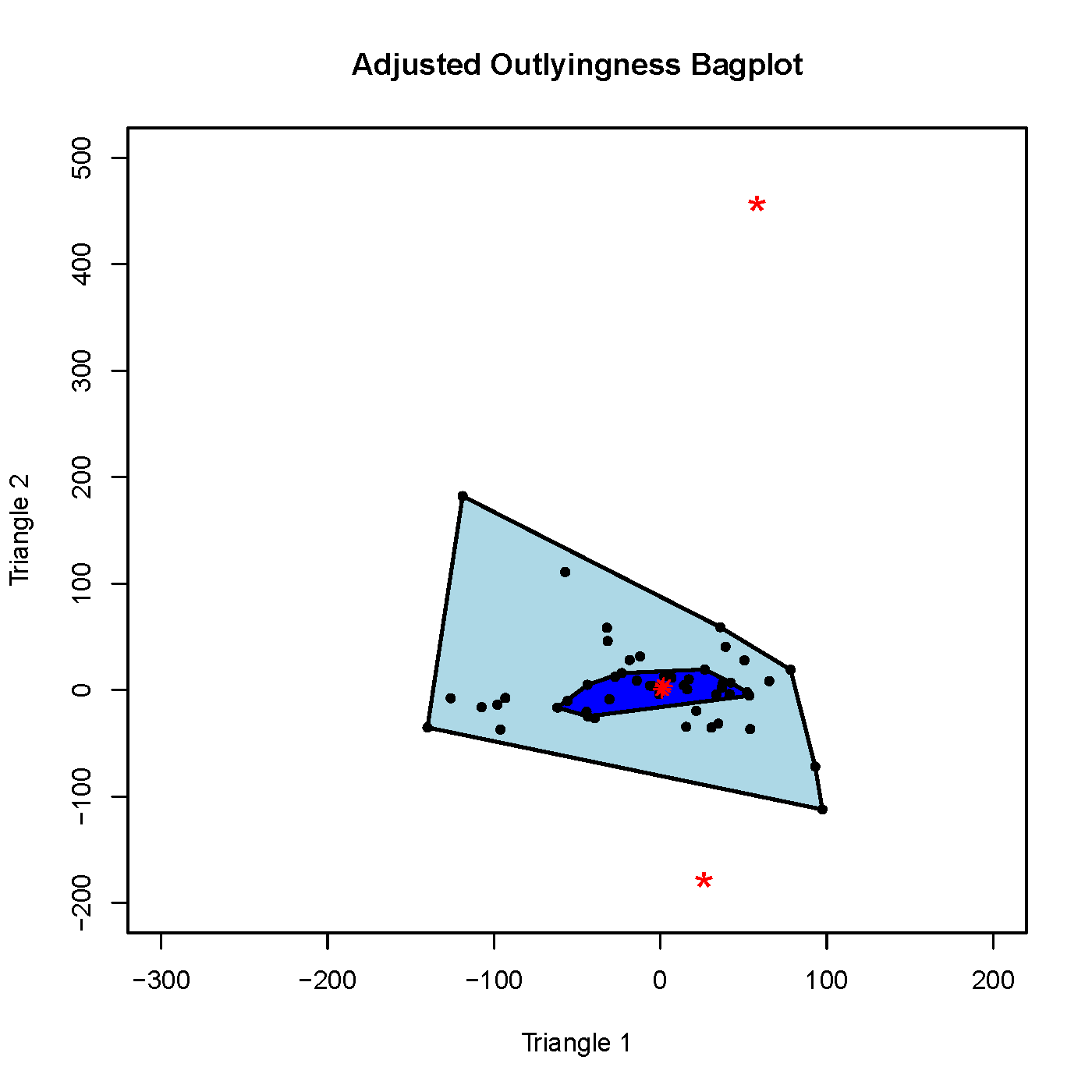}
 	        \caption{Adjusted-Outlyingness Bagplot Using mrfDepth Cut-off Value (Approach 2)}
 	\label{ShBaMe12AOmrfbp}
\end{subfigure}
		\caption{Adjusted-Outlyingness Bagplots (Approaches 1--2)}
		\label{ShBaMe12AOBagplots}
	\end{figure}

\subsubsection{Adjusted Outlyingness} \label{S_AOidentif}
We now explore Adjusted Outlyingness (AO) as an outlier detection and adjustment technique. AO explicitly incorporates a robust measure of skewness and provides a measure of outlyingness for each multivariate observation, alleviating the aforementioned issue with the bagplot based on halfspace depth. Additionally, it may be better equipped to handle more extreme levels of skewness. A bagplot based on AO is also available in the two dimensional case as shown in Figures \ref{ShBaMe12AObp} \rev{and} \ref{ShBaMe12AOBagplots}, which illustrate all three methods outlined in Section \ref{adjoutdescription}.

For these plots, the red asterisk is the point with the lowest AO, and represents a central point of the data analogous to the Tukey median, the dark blue area represents the bag which contains the 50\% of points with the lowest AO and the light blue area represents the loop whose perimeter is constructed by the convex hull of all points not declared as outliers when using the traditional cut-off method (Hubert and Van der Veeken, 2008).  

In Figure \ref{ShBaMe12AObp} we have used approach 3 with the traditional cut-off value given in \cite{HuVe08} 
($\text{cut-off}=Q_3+1.5e^{3MC}IQR$)
which incorporates a robust measure of skewness calculated from the whole data set and hence more fully considers the shape of the data to declare outliers, which are shown in red. This cut-off value was calculated to be 5.5813. The light blue area represents the convex hull of all points not declared as outliers under this methodology and may be considered as an AO\rev{-}based loop. 

The fence approach (approach 1) only captures the shape of the data from the 50\% of observations determined to be least outlying. Figure \ref{ShBaMe12AObpfence} shows the AO bagplot with the fence drawn and we see that under this approach an additional 6 observations are detected as outliers. A further issue that presents itself in this situation is the adjustment to the fence or the loop. 

Finally, we may use the alternative cut-off value as given in the mrfDepth package \citep{mrfDepth} (Approach 2). The cut-off value under this approach is calculated to be 7.6227 compared with 5.5813 under the traditional calculation. An AO bagplot for this approach is shown in Figure \ref{ShBaMe12AOmrfbp} where no fence is drawn. In this case we detect 1 less outlier in comparison to the traditional case. Based on our previous arguments regarding the potential lack of total consideration of skewness under the two alternative approaches we recommend use of the traditional cut-off value (Approach 3).

\begin{figure}[htb]
	\begin{center}
		\includegraphics[width=0.6\textwidth]{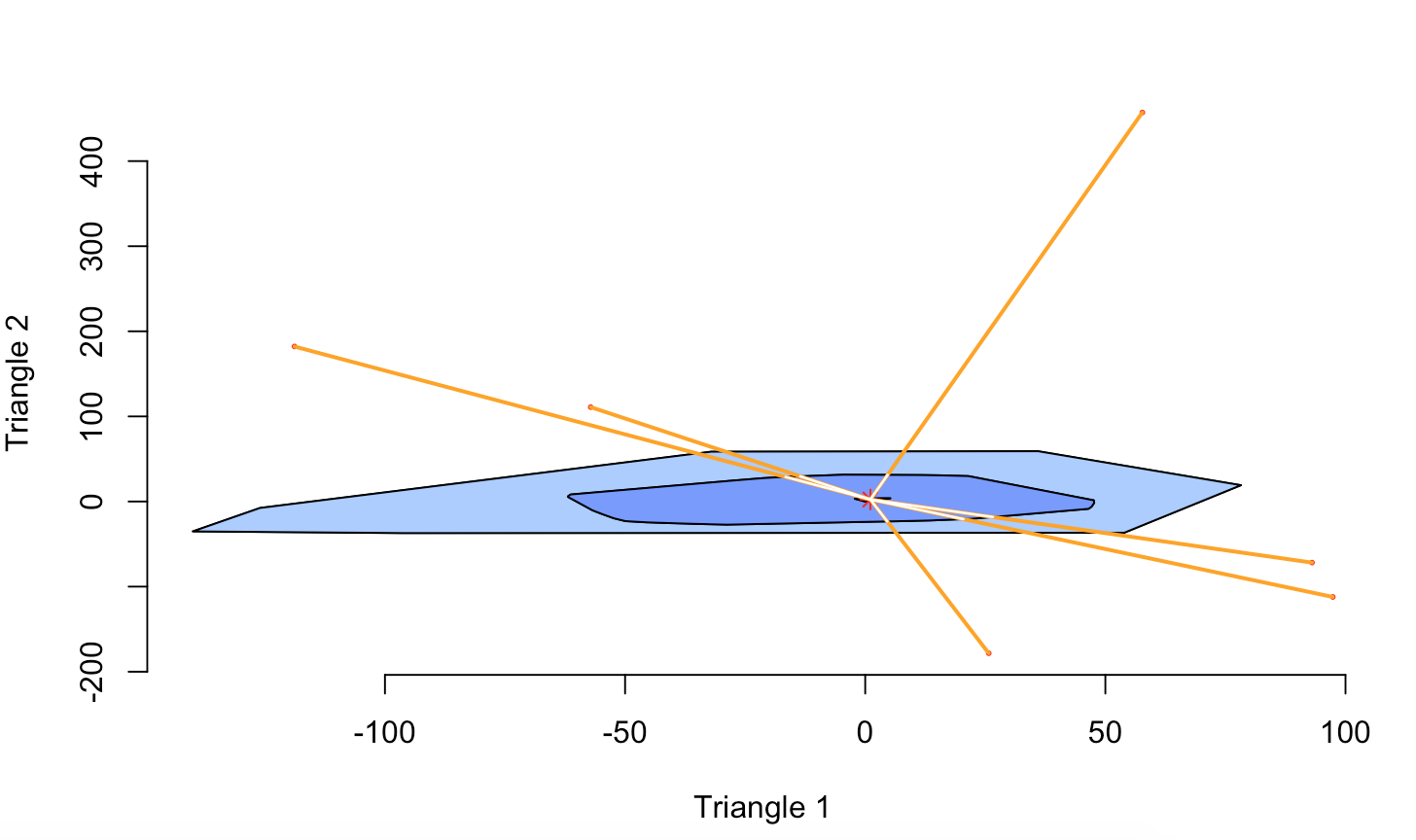}
		\caption{Bagdistance Illustration} 
		\label{ShBaMe12bagdistance}
	\end{center}
\end{figure}

\subsubsection{bagdistance}
The second outlier detection approach we propose is the \textit{bagdistance} (\textit{bd}) (see Section \ref{bagd}) which utilises the bag to capture the shape of the data and provides a distance measure for each observation to represent its outlyingness. A cut-off point is set such that observations with a \textit{bd} beyond this threshold are classified as outliers and are adjusted back to an appropriate point on the ray emanating from \textbf{T}$^*$ passing through \textbf{x}.
An illustration of the \textit{bd} is given in Figure \ref{ShBaMe12bagdistance} where the \textit{bd} for the outliers is given by taking the ratio of the orange line to the white line (noting that the orange line continues through to \textbf{T}$^*$).  
If we choose the \textit{bd} cut-off distance to be uniformly the same as the fence factor for a corresponding bagplot we will detect the same outliers under the two approaches.

\subsubsection{Summary of Detection Results}

Table \ref{ShBaMe12detectiontable} summarises the observations that were detected as outliers by the four techniques discussed here where a tick indicates that the observation was flagged under the relevant technique and a cross indicates that it was not. All observations not listed here were not flagged by any of the techniques. Further, the results in brackets correspond to the halfspace depth, MCD Mahalanobis distance, AO and \textit{bd} values for each observation respectively.

\begin{table}[htb]
	\centering
			\begin{tabular}{|c| c c c c|} \hline
				Outliers &Bagplot &MCD & AO* & \textit{bagdistance}** \\ \hline
				$X_{6,5}$& \checkmark (2) &\checkmark (11.4205)&X (3.8184)  & \checkmark (\rev{3.5744})\\
				$X_{7,3}$ &\checkmark (3)&\checkmark (17.3909)&X (3.7012) & \checkmark (\rev{3.9989})\\
				$X_{7,4}$&\checkmark (1)&\checkmark (22.3643) & X (5.5598) & \checkmark (\rev{4.9284})\\
				$X_{8,2}$&\checkmark (1) &\checkmark (50.4314) &\checkmark (6.2001) & \checkmark (\rev{6.8217})\\ 
				$X_{8,3}$&\checkmark (1)&\checkmark (43.4263)&\checkmark (7.9189) &\checkmark (\rev{7.0153})\\ 
				$X_{9,1}$&\checkmark(1)&\checkmark (262.9189)&\checkmark (15.7057)  &\checkmark (\rev{15.7549}) \\
				$X_{9,2}$ & X (1) &\checkmark (7.6540) &X (1.6197) &X (\rev{4.4161})\\\hline
			\end{tabular}
			\caption{Outlier Detection Results}
			\label{ShBaMe12detectiontable}
		\noindent{\footnotesize  *As detected by Approach 3 (traditional AO cut-off value). **Using cut-off distance of $\sqrt{\chi^2_{0.99,2}}$.}
	\end{table}

\subsection{Adjustment of Outliers}

	In the previous section we outlined how outliers may be detected. Of course, an equally crucial task is to determine how to treat those outliers. In this section we discuss different methods for such treatment.
	
In some robust analyses, if a data point is extremely outlying, it is simply removed. Unfortunately, this option is not available in most triangular reserving techniques (which require a fully populated triangle), and hence we must formulate and choose appropriate adjustment mechanisms. 

			\begin{figure}[htb]
		\centering 
		\begin{subfigure}{.45\textwidth}
			\centering
			\includegraphics[width=\textwidth]{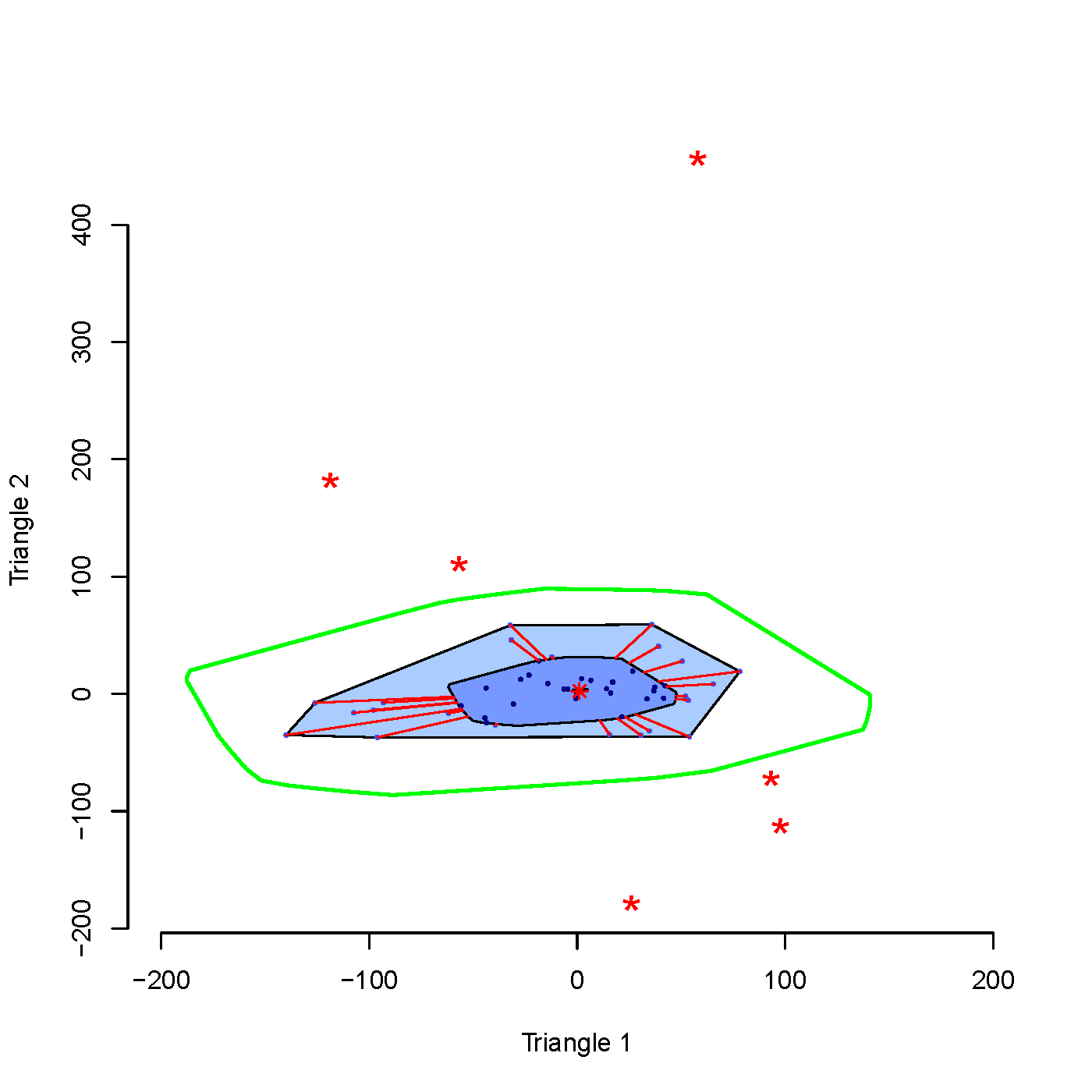}
			\caption{\scriptsize Bagplot before Adjusting Outliers with Fence Drawn}
			\label{ShBaMe12bpfence}
		\end{subfigure}%
		\begin{subfigure}{.45\textwidth}
			\centering			 	 		\includegraphics[width=\textwidth]{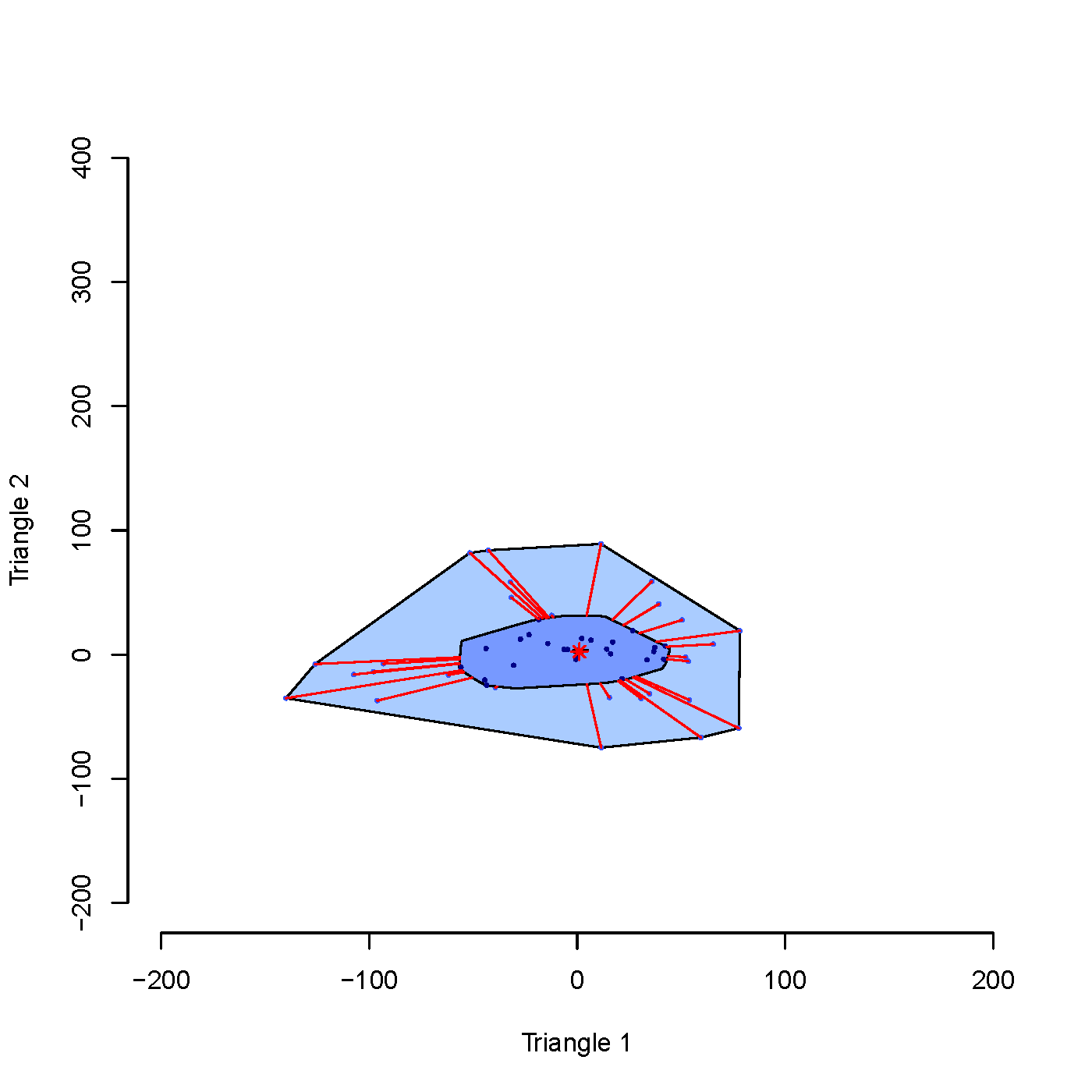}
			\caption{\scriptsize Bagplot After Adjusting Outliers to Fence and Subsequently Recalculating the Loop}
			\label{ShBaMe12bpadjfence}
		\end{subfigure}
		\caption{}
		\label{ShBaMe12BPadjfence}
	\end{figure}
	
\subsubsection{Bagplot Based Adjustments}
	
Under the bagplot approach, \cite{VeVa11} suggest that outlying observations should be brought back to the fence however upon inspecting their plots they appear to have brought observations back to the loop.

Figure \ref{ShBaMe12bpfence} and Figure \ref{ShBaMe12bpadjfence} show the bagplot with the fence drawn in green and the bagplot after adjusting residuals back to the fence respectively. In Figure \ref{ShBaMe12bpadjfence},  all of the adjusted outliers are within the loop, however, this is not always the case as  after adjustment of the outlying observations, the bagplot methodology is performed again and with some data sets, some of the observations will be outlying because the overall shape of the data has changed which will lead to a different bag and hence fence. 

\begin{figure}[htb]
	\begin{center}
		\includegraphics[width=0.5\textwidth]{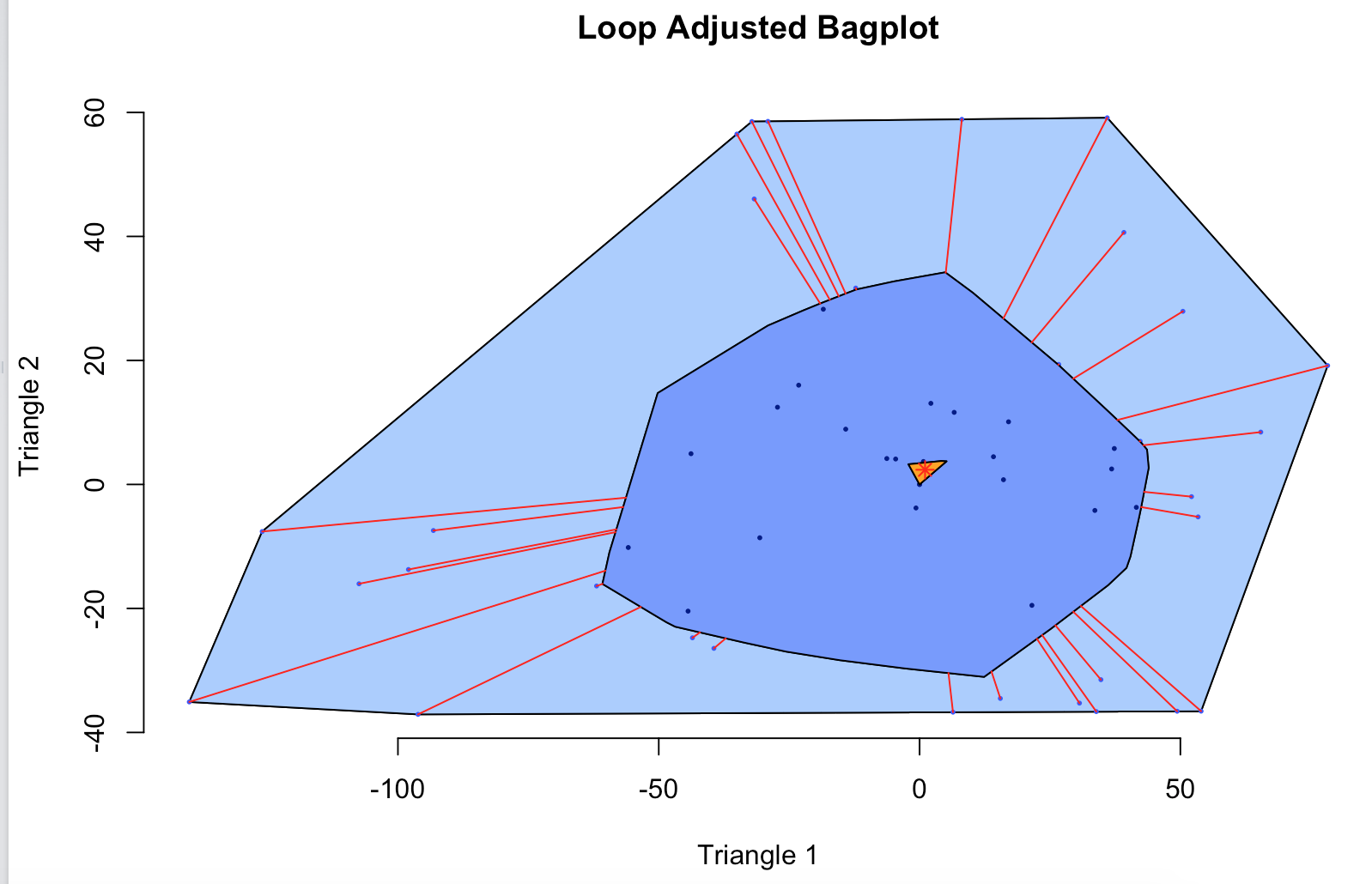}
		\caption{Bagplot After Adjusting Outliers to Loop} 
		\label{ShBaMe12bpadjloop}
	\end{center}
\end{figure}

Figure \ref{ShBaMe12bpadjloop} shows the bagplot after adjusting outliers to the loop. Note that the adjustment in this case is more drastic in comparison to the fence adjustment. 

\subsubsection{MCD Mahalanobis Distance Based Adjustments}
The adjustment methodology under the MCD Mahalanobis Distance approach is outlined in Section \ref{MCD}.

	
\subsubsection{Bagdistance Based Adjustments}\label{bdadjsec}

	Through calculation of the \textit{bd} for each observation a similar adjustment technique as was used in the MCD Mahalanobis example can be employed in that we adjust an outlying observation \textbf{x} according to 
	\begin{equation}
	\text{min}\left(\frac{f}{\textit{bd}},1\right)(\textbf{\text{x}}-\textbf{T}^*)+\textbf{T}^*,
	\end{equation}
	where $f$ represents the factor of the bag that we wish to adjust outliers back to. For adjustment back to the bagplot fence we choose $f=\sqrt{\chi^2_{0.99,2}}$. Note that other adjustment functions may be employed. 
	
	For illustration purposes, we have considered two other adjustment functions such that, dependant on the level of outlyingness as measured by \textit{bd}, observations are adjusted to differing degrees. 
	
	In the first case all observations (\textbf{x}) are adjusted according to 
	\begin{equation}
	\textbf{x}^{rob}=\begin{cases}
	\textbf{x}, & \text{if } \textit{bd}_\textbf{x}\leq f; \\
	\text{min}\left(\frac{f}{\textit{bd}_\textbf{x}},1\right)(\textbf{\text{x}}-\textbf{T}^*)+\textbf{T}^*, & \text{if } f<\textit{bd}_\textbf{x}\leq\frac{1}{2}\left(2f+\sqrt{4f+1}+1\right);\\
	\frac{f+\sqrt{\textit{bd}_\textbf{x}}}{\textit{bd}_\textbf{x}}(\textbf{\text{x}}-\textbf{T}^*)+\textbf{T}^*, & \text{if } \textit{bd}_\textbf{x}>\frac{1}{2}\left(2f+\sqrt{4f+1}+1\right).
	\end{cases}\label{bdadj1}
	\end{equation}
	This means that moderate outliers will be brought back to the fence whereas more extreme outliers will be adjusted to points beyond the fence according to their levels of outlyingness. \rev{Specifically, the $\frac{1}{2}(2f+\sqrt{4f+1}+1)$ constraint is chosen as it is the value of $\textit{bd}_\textbf{x}$ where the adjustment function in case 3 of equations \ref{bdadj1} and \ref{bdadj2} (i.e. $\frac{f+\sqrt{\textit{bd}_\textbf{x}}}{\textit{bd}_\textbf{x}}$) is equal to 1. For $\textit{bd}_\textbf{x}>\frac{1}{2}\left(2f+\sqrt{4f+1}+1\right)$, $\frac{f}{\textit{bd}_\textbf{x}}<\frac{f+\sqrt{\textit{bd}_\textbf{x}}}{\textit{bd}_\textbf{x}}<1$ which means that the points that fall into case 3 will be adjusted back towards the centre of the data, however, to a point beyond the fence. For completeness, $\frac{f+\sqrt{\textit{bd}_\textbf{x}}}{\textit{bd}_\textbf{x}}>1$ for $\textit{bd}_\textbf{x}<\frac{1}{2}\left(2f+\sqrt{4f+1}+1\right)$,    which means that if the constraint was set at a lower value, then case 3 could lead to outliers being adjusted further away from the central point of the data rather than closer to it.}This takes one view point on outliers whereby if an observation is very far outlying then it should still represent  moderately rare events in the data. Another approach would be to consider outliers beyond a certain limit as misleading or errors and adjust them further towards the centre of the data, limiting their influence. Further investigation will be required based on the data set being analysed and the adjustment functions can then be designed depending on the conclusions drawn. 
	
	The second approach we consider is given by 
	\begin{equation}
	\textbf{x}^{rob}=\begin{cases}
	\textbf{x}, & \text{if } \textit{bd}_\textbf{x}\leq f; \\
	\text{min}\left(\frac{f}{\textit{bd}_\textbf{x}},1\right)(\textbf{\text{x}}-\textbf{T}^*)+\textbf{T}^*, & \text{if } f<\textit{bd}_\textbf{x}\leq\frac{1}{2}\left(2f+\sqrt{4f+1}+1\right)\text{ OR } \textit{bd}_\textbf{x}>u\;\rev{>f};\\
	\frac{f+\sqrt{\textit{bd}_\textbf{x}}}{\textit{bd}_\textbf{x}}(\textbf{\text{x}}-\textbf{T}^*)+\textbf{T}^*, & \text{if }\frac{1}{2}\left(2f+\sqrt{4f+1}+1\right)<\textit{bd}_\textbf{x}\;\rev{\leq u}.
	\end{cases}\label{bdadj2}
	\end{equation} 
	Under this adjustment function, we set an additional limit $u$ such that if outliers are beyond this point they are also brought back to the fence. The rationale for this approach is that if an observation is this far outlying it should be given full adjustment back to the fence as it either represents a data entry error or the information contained in that observation is too irrelevant for the current task to include more fully. \rev{Note that $u$ is an arbitrary selection that will be dependent on the data set under review. The key takeaway is that with unique measures of outlyingness for each observation, there is significantly greater flexibility in how observations can be adjusted.}
	
			\begin{figure}[htb]
			\centering 
			\begin{subfigure}{.5\textwidth}
				\centering
				\includegraphics[width=\textwidth]{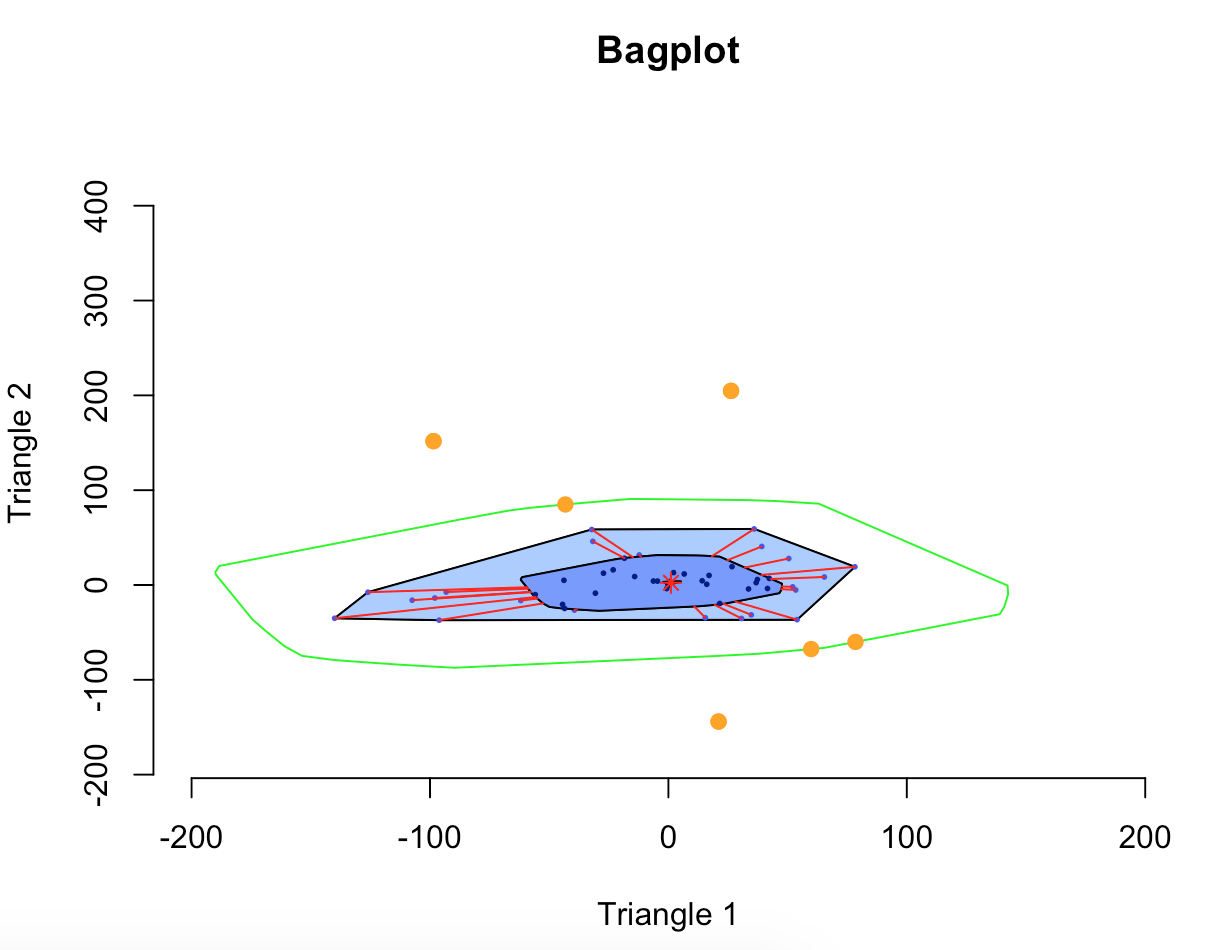}
				\caption{\scriptsize Outliers Adjusted According to Equation \eqref{bdadj1}}
				\label{bdadj1plot}
			\end{subfigure}%
			\begin{subfigure}{.5\textwidth}
				\centering
				\includegraphics[width=\textwidth]{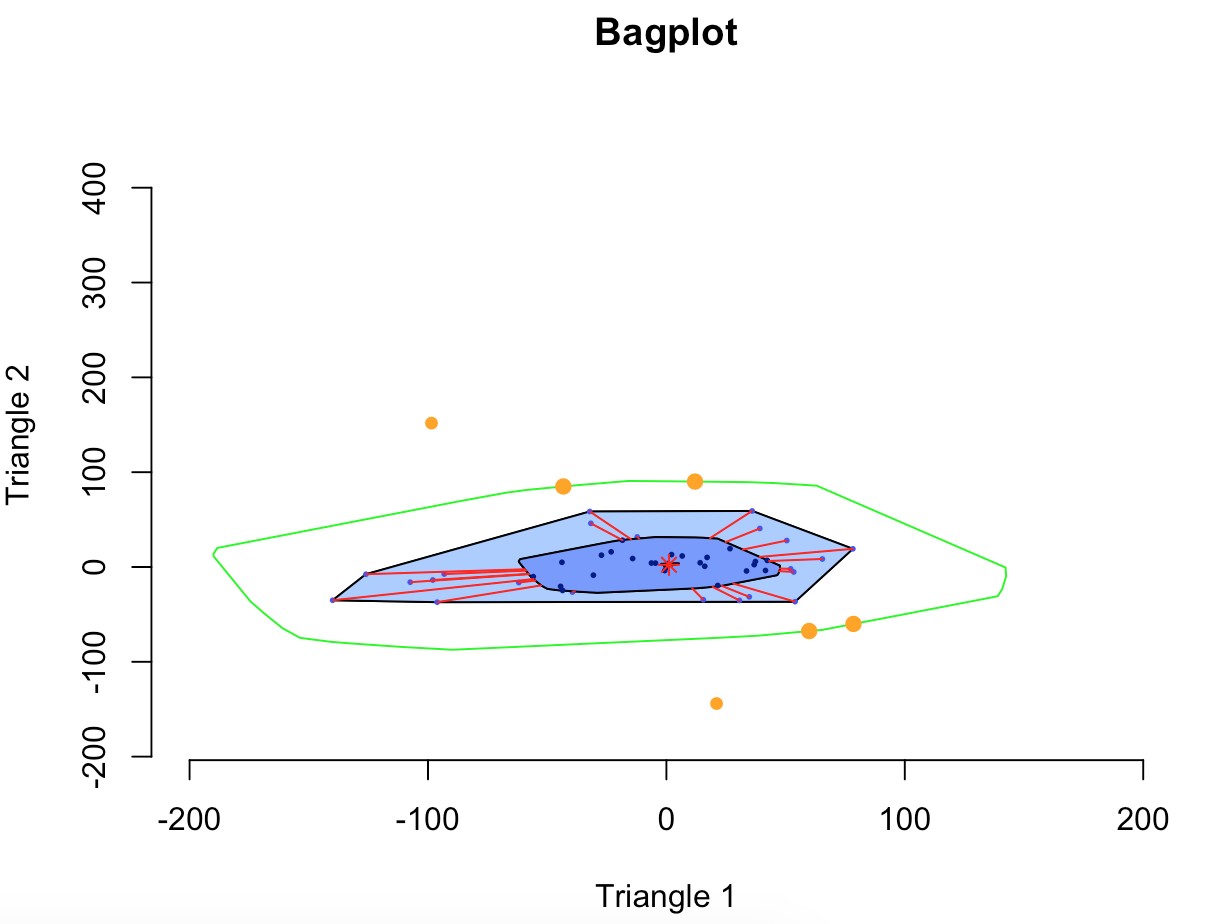}
				\caption{\scriptsize Outliers Adjusted According to Equation \eqref{bdadj2}}
				\label{bdadj2plot}
			\end{subfigure}
			\caption{Bagdistance Adjusted Bagplots}
			\label{bdadj}
		\end{figure}
		
	Figure \ref{bdadj1plot} shows the outliers after adjustment according to Equation \eqref{bdadj1} and Figure \ref{bdadj2plot} shows the outliers after adjustment according to Equation \eqref{bdadj2} where we have set $f=\sqrt{\chi^2_{0.99,2}}$ and $u=15$. \rev{Under the latter approach with $u=15$, observation $X_{9,1}$ which has bagdistance of 15.8 as shown in Table \ref{ShBaMe12detectiontable} is brought back to the fence whereas it remains beyond the fence in Figure \ref{bdadj1plot}.} 

 \begin{remark}
Additionally, we may use the \textit{bd} values as a residual term when fitting a model with a robust loss function (e.g. robust M-estimation). For example we could utilise the \textit{bd} in Huber's loss function such that we have \begin{equation}
	L_c^{\textit{bd}}(\textbf{x})=\begin{cases}
	\frac{1}{2}\textit{bd}^2, & \text{if } \textit{bd}\leq c; \\ 
	c(|\textit{bd}|-\frac{1}{2}c) & \text{otherwise}. 
	\end{cases}
	\end{equation}
 \end{remark}

\subsubsection{Adjusted Outlyingness Based Adjustments}
	
The AO approach provides a measure of outlyingness for each observation. This measure is based on numerous one-dimensional projections of the individual multivariate observations and the total multivariate sample. For each of these projections, the univariate measure of AO is calculated. The maximum of these univariate AO values after all the projections is taken to be the AO measure in the multivariate case. Without knowing which direction has led to the final AO we are seemingly unable to backtransform the AO measure to lead to adjusted residuals and ultimately adjusted claim amounts. A solution is to use the AO-based bagplot such that we adjust outliers back to the fence or loop. Consideration must be given to which cut-off value or whether a fence has been drawn to detect outliers under the AO approach. In each case the loop may be different. We recommend using the traditional cut-off value (Approach 3) as it explicitly incorporates skewness.  

\subsubsection{Summary of Adjustment Mechanisms}

				Table\rev{s} \ref{ShBaMe12adjustmenttable1} and \ref{ShBaMe12adjustmenttable2} summarise the claim values for the outliers detected and their adjusted values for Triangle 1 and 2 respectively. Note that in the case of the AO-Fence and AO-Loop adjustments we are only considering those observations that were flagged as outliers based on the traditional cut-off. 
				\begin{table}[htb]
					\begin{center}
						\resizebox{1\linewidth}{!}{%
							\begin{tabular}{|c| c | c| c| c| c|c|c|c|c|} \hline
								Outliers & Initial & MCD  &BP-Fence &BP-Loop  & AO-Fence & AO-Loop& AO-mrfDepth&\textit{bd}-\eqref{bdadj1} (no limit) &\textit{bd}-\eqref{bdadj2} (limit) \\ \hline
								$X_{6,5}$ &  238 375 &  226 990 &  \rev{231 580} &  \rev{219 269} & - & - & - &  \rev{232 213} &  \rev{232 213} \\ 
								$X_{7,3}$ &  633 500 &  652 977 &  \rev{645 490} &  \rev{656 826} & - & - & - &  \rev{645 081} &  \rev{645 081} \\ 
								$X_{7,4}$ &  432 257 &  403 580 &  \rev{409 118} &  \rev{393 805} & - & - & - &  \rev{409 668} &  \rev{409 668} \\ 
								$X_{8,2}$ & 1 458 541 & 1 557 335 & \rev{1 543 798} & \rev{1 565 203} & 1 802 824 & 1 588 895 & - & \rev{1 484 742} & \rev{1 484 742} \\ 
								$X_{8,3}$ &  727 098 &  713 520 &  \rev{715 180} &  \rev{710 780} & 702 174 &  700 640 &  700 177 &  \rev{723 168} &  \rev{723 168} \\ 
								$X_{9,1}$ & 2 210 754 & 2 139 306 & \rev{2 143 332} & \rev{2 137 794} & 2 119 751 & 2 133 590 & 2 133 589 & \rev{2 164 880} & \rev{2 144 073} \\ 
								$X_{9,2}$ & 1 517 501 & 1 538 538 & - & - & - & - & - & - & - \\ 
								\hline
							\end{tabular}}
							\caption{Triangle 1 Outlier Adjustment Results}
							\label{ShBaMe12adjustmenttable1}
						\end{center}
					
						\begin{center}
							\resizebox{1\linewidth}{!}{%
								\begin{tabular}{|c|c| c| c| c| c| c |c|c |c|} \hline
									Outliers & Initial & MCD  &BP-Fence &BP-Loop  & AO-Fence & AO-Loop & AO-mrfDepth&\textit{bd}-\eqref{bdadj1} (no limit) &\textit{bd}-\eqref{bdadj2} (limit) \\ \hline
									$X_{6,5}$ & 6 650 & 9 124 & \rev{8 215} & 11 050 & - & - & - & \rev{8 049} & \rev{8 049} \\ 
									$X_{7,3}$ & 86 734 & 75 546 & \rev{80 162} & 73 947 & - & - & - & \rev{80 344} & \rev{80 344} \\ 
									$X_{7,4}$ & 18 109 & 29 028 & \rev{27 279} & 33 348 & - & - & - & \rev{26 990} & \rev{26 990} \\ 
									$X_{8,2}$ & 132 208 & 98 329 & \rev{103 742} & \rev{96 595} & 17 257 & 88 685 & - & \rev{123 419} & \rev{123 419} \\ 
									$X_{8,3}$ & 20 923 & 49 957 & \rev{47 709} & 57 597 & 76 939 & 80 387 & 81 427 & \rev{29 809} & \rev{29 809} \\ 
									$X_{9,1}$ & 112 103 & 43 470 & \rev{47 024} & \rev{41 679} & 24 263 & 37 621 & 37 620 & \rev{67 438} & \rev{47 179} \\ 
									$X_{9,2}$ & 33 250 & 34 062 & - & - & - & - & - & - & - \\ 
									\hline
								\end{tabular}}
								\caption{Triangle 2 Outlier Adjustment Results}
								\label{ShBaMe12adjustmenttable2}
							\end{center}
						\end{table}

For the AO-mrfDepth approach we are considering outliers flagged under this methodology and outliers are adjusted back to the loop in this case. As we adjust residuals back in the direction of the central estimate of the data this may lead to an upward or downward adjustment for each residual and similar adjustments will be seen for the corresponding claim observations. Note that '-' values are provided where outliers were not detected under that methodology.

\subsection{Final Reserves and Discussion}

Once the outliers have been appropriately identified and treated, the data can be used to compute reserves estimates. Here we used the multivariate time series chain-ladder technique as described in \cite{MeWu08}.

\begin{table}[htb]
						\begin{center}
							\resizebox{\linewidth}{!}{
								\begin{tabular}{|c|c|c|c|c|c|c|c|c|c|} \hline
									\multirow{2}{*}{} & \multirow{2}{70pt}{\centering \rev{No. of Outliers Detected}} &\multicolumn{2}{c|}{Triangle 1}& \multicolumn{2}{c|}{Triangle 2}&\multicolumn{2}{c|}{Total}&\multicolumn{2}{c|}{\rev{Difference (\%)}} \\ \cline{3-10}
									& & Reserve & \rev{RMSE} & Reserve & \rev{RMSE} & Reserve & \rev{RMSE} & \rev{Reserve} & \rev{RMSE} \\ \hline										
									Original & \rev{--} & 6 435 951 & 322 573 & 489 028 & 90 542 & \textbf{6 924 978} & \textbf{337 001} & \rev{\textbf{--}} & \rev{\textbf{--}} \\ 
									MCD & \rev{7} & 6 438 541 & 283 054 & 438 497 & 45 552 & \textbf{6 877 037} & \textbf{293 870} & \rev{\textbf{-0.69}} & \rev{\textbf{-12.80}} \\ 
									Bagplot-Fence & \rev{6} & \rev{6 427 705} & \rev{291 416} & \rev{441 754} & \rev{50 822} & \textbf{\rev{6 869 460}} & \textbf{\rev{301 923}} & \rev{\textbf{-0.80}} & \rev{\textbf{-10.41}} \\ 
									Bagplot-Loop & \rev{6} &\rev{6 416 549} & \rev{283 191} & \rev{445 097} & \rev{41 669} & \textbf{\rev{6 861 647}} & \textbf{\rev{295 999}} & \rev{\textbf{-0.91}} & \rev{\textbf{-12.17}} \\ 
									AO-Fence & \rev{3} & 6 509 807 & 294 572 & 389 602 & 62 491 & \textbf{6 899 409} & \textbf{288 668} & \rev{\textbf{-0.37}} & \rev{\textbf{-14.34}} \\ 
									AO-Loop & \rev{3} & 6 456 446 & 292 021 & 444 764 & 47 660 & \textbf{6 901 210} & \textbf{297 914}  & \rev{\textbf{-0.34}} & \rev{\textbf{-11.60}} \\ 
									AO-mrfDepth & \rev{2} & \rev{6 456 381} & \rev{291 999} & \rev{445 496} & \rev{47 722} & \textbf{\rev{6 901 876}} & \textbf{\rev{297 901}} & \rev{\textbf{-0.33}} & \rev{\textbf{-11.60}} \\ 
									\textit{bd}-\eqref{bdadj1} (no limit) & \rev{6} & \rev{6 417 482} & \rev{304 232} & \rev{456 439} & \rev{68 329} & \textbf{\rev{6 873 921}} & \textbf{\rev{315 131}} & \rev{\textbf{-0.74}} & \rev{\textbf{-6.49}} \\ 
									\textit{bd}-\eqref{bdadj2} (limit) & \rev{6} & \rev{6 412 956} & \rev{301 127} & \rev{438 432} & \rev{61 425} & \textbf{\rev{6 851 388}} & \textbf{\rev{306 010}} & \rev{\textbf{-1.07}} & \rev{\textbf{-9.20}} \\ 
									\hline
								\end{tabular}}
								\caption{Bivariate Example Reserves}
								\label{ShBaMe12reservestable}
							\end{center}
						\end{table}
						
Table \ref{ShBaMe12reservestable} summarises the final reserve estimates and their associated \rev{RMSE} for each individual triangle and total reserves under each outlier detection and adjustment technique as well as when we simply apply the multivariate chain-ladder without adjusting any observations.  However we consider the last three development periods as separate univariate triangles. This is because there are few data points for these development periods and applying the multivariate chain-ladder to such periods often leads to highly volatile results or potentially failure in that elements of the estimated correlation matrices may have absolute values greater than one. This in turn may lead to a lack of convergence. Note that some authors \citep[including][]{MeWu08} suggest extrapolation of these correlation variables from the previous periods however as our main focus is on outlier detection and adjustment we have not pursued this option.
					
The robust reserves in this example are always modestly less than the original reserves. This phenomenon is well known in the robust literature, and is obviously related to all adjustements being \emph{reductions}; losses are thus decreased and so are the associated calculated reserves. The greatest adjustment is for the \textit{bagdistance} technique where a limit is set for the adjustment function. 

In each case, we saw a much greater reduction in the \rev{RMSE} than reserves. For example, the bagplot-loop methodology leads to a \rev{0.91}\% reduction in reserves and a \rev{12.17}\% reduction in \rev{RMSE}. This suggests that even if the change in reserve estimates are only minor, the accuracy of such estimates may be enhanced considerably as a result of these robust techniques. 

\section{Detection and Treatment of Outliers in N-Dimensional Reserving} \label{Ndimframe}

\subsection{An N-dimensional Framework}
We now put forward a framework for N-dimensional robust chain-ladder reserving. The methodology is as follows. Firstly, perform the robust Poisson GLM chain-ladder technique on each triangle separately to obtain residuals. These residuals are then stored in an N-dimensional matrix. It is these N-dimensional residuals that outlier detection and adjustment techniques are applied to. The techniques are based on AO, halfspace depth, MCD Mahalanobis distance and \textit{bagdistance}. 

Under the AO approach outliers are detected using the standard cut-off value (see Section \ref{adjoutdescription}). In the three dimensional case a graphical representation is also constructed. Firstly, declare the point with the lowest AO as the AO median. Next, form the convex polyhedron that contains the 50\% of points with the lowest AO which can be considered as a N-dimensional AO based bag. We then construct an N-dimensional AO based loop by forming the convex polyhedron that contains all non-outlying points.

 Outliers are adjusted by bringing them back to the intersection point of the ray from the AO-median to the outlying point and the AO based loop. This intersection is found using the parametric line clipping algorithm \citep{CyBe78,LiBa84}. Adjusted residuals are then backtransformed to give robust incremental claims. The multivariate time series chain-ladder \citep{MeWu08} is applied on these robust claims. 
 
 For the halfspace depth approach we calculate residuals in the same fashion, however the outlier detection and treatment methodology is altered. The halfspace depth methodology outlined in Section \ref{hddescription} can be applied in N dimensions with Figure \ref{halfspace} providing an illustration in 2 dimensions. Firstly, calculate the halfspace depth of each observation.
 The Tukey Median is then the observation with the greatest halfspace depth or the centre of gravity of the deepest region if this is not a unique point. Next, formulate the N-dimensional bag. Firstly, let $D_s$ represent the region that contains all points with halfspace depth greater than or equal to $s$. Let $\#D_s$ be the number of data points in $D_s$. The bag is constructed by linearly interpolating the two convex regions that satisfy $\#D_s\leq \lfloor \frac{n}{2} \rfloor <\#D_{s-1}$  with respect to the Tukey Median. This interpolation is done in the same way as for the bivariate bagplot as described in \citet{MiRaRoSeSoStSt03} however now applicable for any polytype rather than only two dimensional polygons. Firstly, calculate 
 \begin{equation}
     \lambda=\frac{\frac{n}{2}-\#D_s}{\#D_{s-1}-\#D_s}.
 \end{equation}
 The bag is given by $\lambda\cdot\text{outer polytype}+(1-\lambda)\cdot\text{inner polytype}$. Next, the N-dimensional fence is constructed by multiplying the N-dimensional bag by a fence factor with respect to the Tukey Median. Outliers are declared as points beyond the fence. The fence factor we use is $\sqrt{\chi^2_N}$. For all $N>1$, this is greater than the previously used arbitrary value of 3 in the two dimensional case and hence less outliers will be detected and adjusted.
Then construct the N-dimensional loop by forming the convex hull of all non-outlying points.
Outliers are treated by bringing them back to either the fence or loop. In three dimensions, the bag, loop and fence are convex polyhedrons and graphical representation is available. Adjusted residuals are backtransformed and the multivariate time series chain-ladder applied.

 The MCD Mahalanobis distance and \textit{bagdistance} approaches are implemented in the same fashion as they were for the bivariate case. 
For the trivariate case a graphical representation of the tolerance ellipsoid for the MCD Mahalanobis technique may be displayed. We now show the implementation of these four techniques on real data and compare the results. 

\subsection{Illustration with the AUSI dataset}\label{trivareg}

For the purposes of illustration we use data from three different lines of business of two Australian insurers. A short discussion of this data is provided in Appendix \ref{tridata}. 

\subsubsection{Data Adjustments}

The fitting of the robust Poisson GLM stage of the methodology is only possible if there are no negative incremental claims in each triangle. However, of the 820 observations for each line, the CTP line from insurer 1, the CTP line from insurer 2 and the Home line from insurer 1 have 1, 21 and 57 negative claims respectively. We have set all these claims to zero. To ensure that this did not significantly impact results, we calculated development factors under the classical chain-ladder technique before and after the adjustment. The largest change in a development factor was 0.231\% and hence the effect of the adjustment is small and is deemed acceptable.

\begin{figure}[H]
	\centering 
	\begin{subfigure}{.49\textwidth}
		\centering
		\includegraphics[width=.9\textwidth]{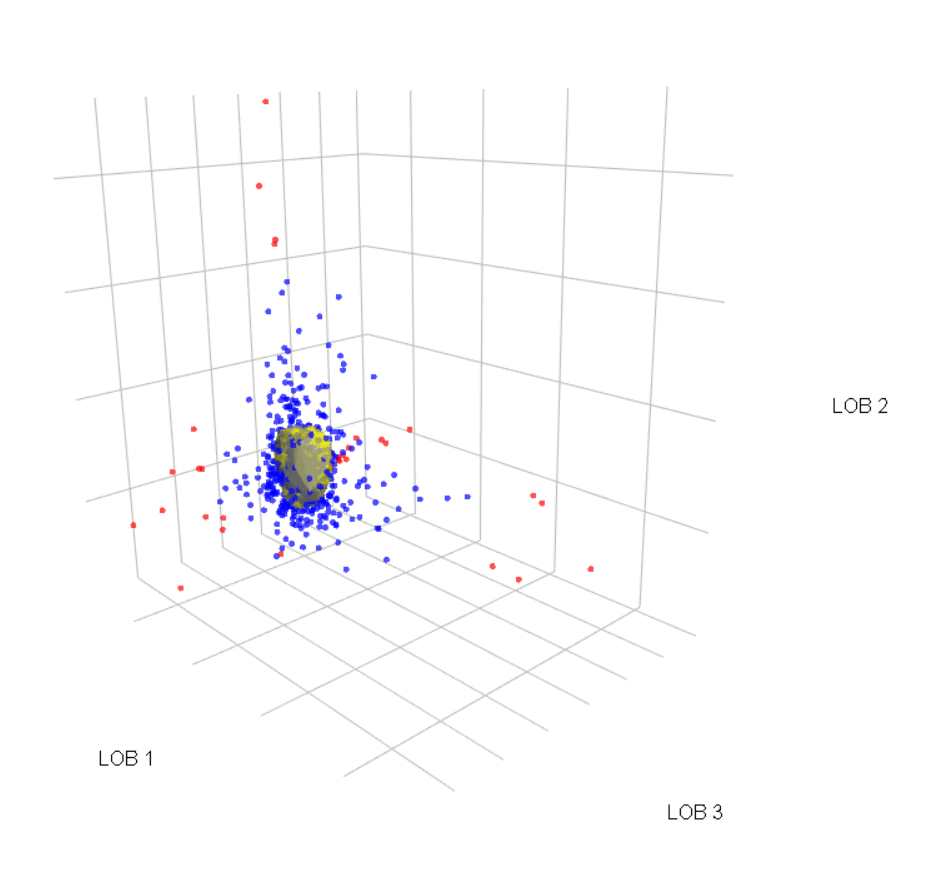}
		\caption{Trivariate Residuals with 3D AO-based Bag}
		\label{trivarbag}
	\end{subfigure}
	\begin{subfigure}{.49\textwidth}
		\centering			 	 		
		\includegraphics[width=.9\textwidth]{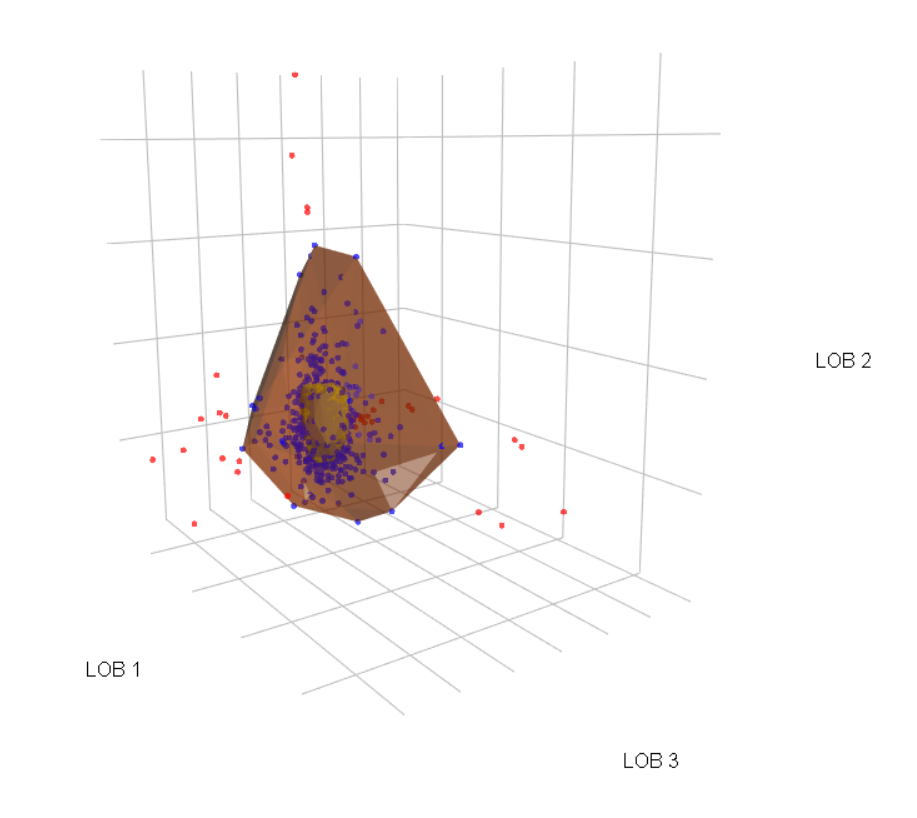}
		\caption{Trivariate Residuals with 3D AO-based Bag and Loop (Approach 3)}
		\label{trivarbagloop}
	\end{subfigure}
	\caption{AO Based Detection}
	\label{trivarbagandfence}
\end{figure}

\subsubsection{Detection of outliers}
Firstly, we perform the AO technique on these residuals. 34 outliers were flagged under this approach. We then construct a 3D version of the AO-based bagplot such that the 50\% of data points with the lowest AO are contained in a convex polyhedron analogous to the bag and an additional convex polyhedron is formed for all non-outlying observations analogous to the loop. 
Figure \ref{trivarbag} shows the residuals with the 3D AO bag drawn in yellow and Figure \ref{trivarbagloop} shows the residuals with the 3D AO loop in orange and the corresponding bag contained inside. Outliers are shown in red in each case.

We note that this loop extends in the various directions of the data's dispersion and seemingly captures its general shape. For completeness we have also investigated the use of the cut-off value for AO of $\sqrt{\chi^2_{\{99,N\}}}\cdot\text{median}(AO)$ as suggested in \citet{mrfDepth}. Under this approach 48 outliers are detected.

\begin{figure}[H]
	\centering 
	\begin{subfigure}{.49\textwidth}
		\centering
		\includegraphics[width=.9\textwidth]{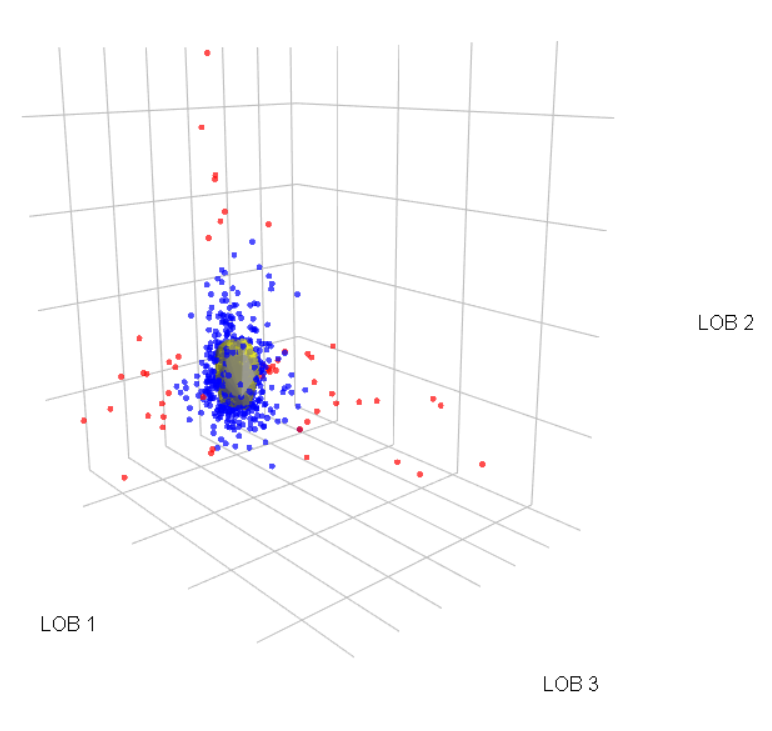}
		\caption{Trivariate Residuals with 3D HD-mrfDepth-based Bag}
		\label{trivarHDmrfDepthbag}
	\end{subfigure}
	\begin{subfigure}{.49\textwidth}
		\centering	\includegraphics[width=.9\textwidth]{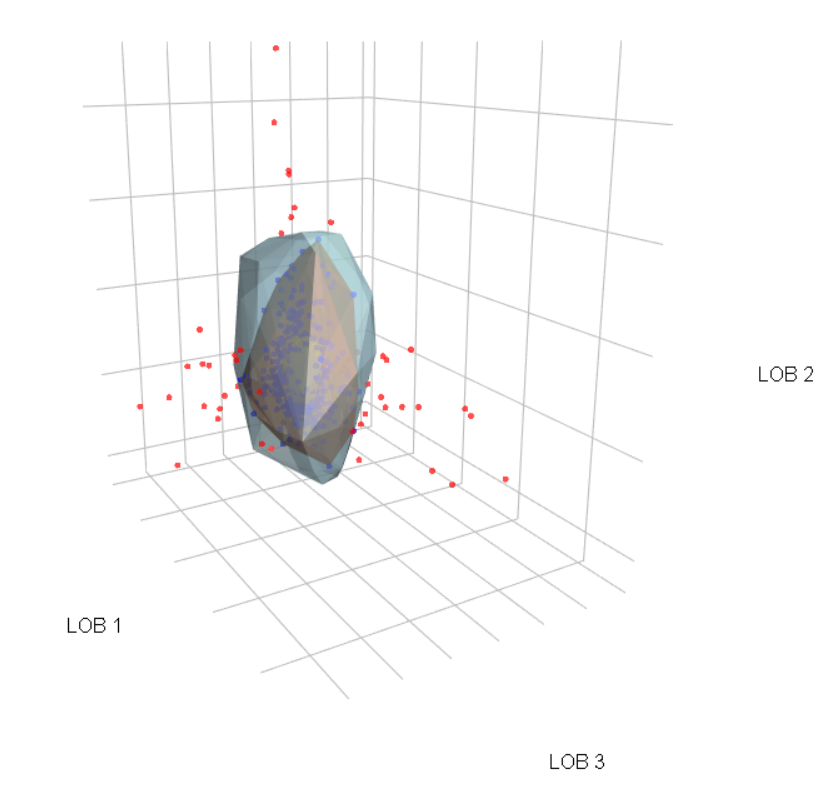}
		\caption{Trivariate Residuals with 3D HD-mrfDepth-based Bag, Loop and Fence}
		\label{trivarHDmrfDepthbagloopfence}
	\end{subfigure}
	\caption{Halfspace Depth Based Detection}
	\label{trivarhdepthmrfbagloopandfence}
\end{figure}

We now illustrate the halfspace depth based approach. Figure \ref{trivarHDmrfDepthbag} shows the residuals with the bag drawn in yellow and Figure \ref{trivarHDmrfDepthbagloopfence} shows the residuals with the bag loop and fence all drawn. The loop is again in orange and is within the light blue fence. Outliers are declared as the points outside the fence and under this methodology we detect \rev{48} outliers. Interestingly, \rev{29} of the 34 observations flagged as outliers under the traditional AO methodology are also detected under the halfspace depth approach. This highlights some agreement between the two methodologies. Note that the bags under both the AO and halfspace depth approaches are relatively similar, however the fence which is used to declare outliers under the halfspace depth methodology appears unable to capture the three dimensional skewness in the data. Rather, it remains relatively elliptical in shape whereas the AO loop stretches in each direction that the data is dispersed. This is further reflected in the number of outliers flagged. We conject that the halfspace depth approach may be misclassifying regular observations as outliers in this instance. 

	\begin{figure}[H]
		\centering 
		\begin{subfigure}{.49\textwidth}
			\centering
			\includegraphics[width=.9\textwidth]{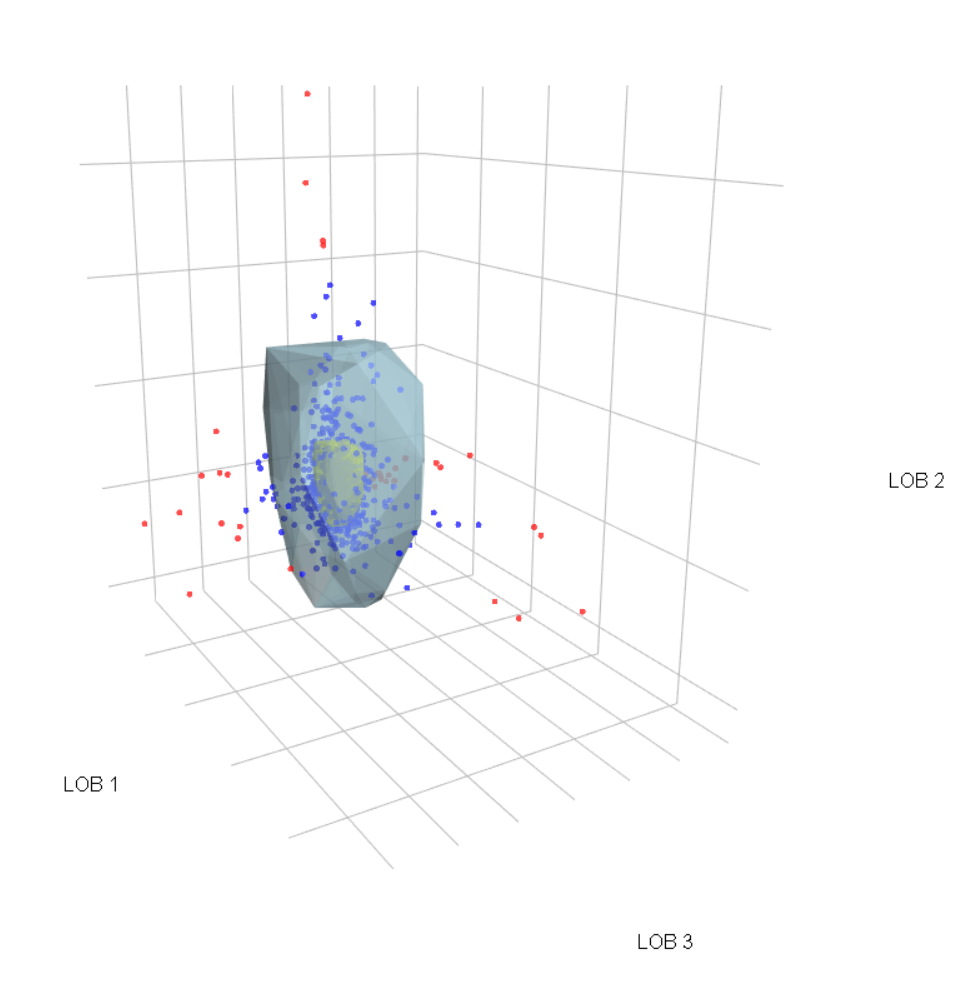}
			\caption{AO Based Bag and Fence (Approach 1)}
			\label{trivarbagfence}
		\end{subfigure}
		\begin{subfigure}{.49\textwidth}
			\centering	\includegraphics[width=.9\textwidth]{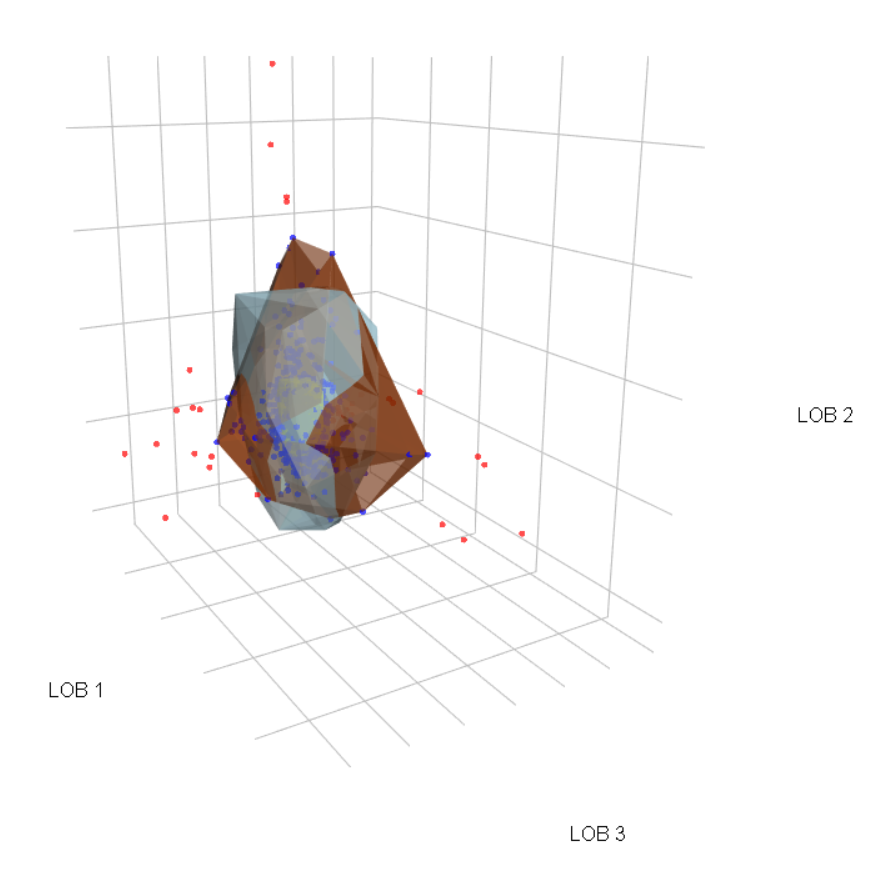}
			\caption{AO Based Bag, Loop and Fence}
			\label{trivarbagloopfence}
		\end{subfigure}
		\caption{AO Based Detection (Approach 1 vs Approach 3)}
		\label{trivarAOfence}
	\end{figure}

 To highlight the differences between a fence-based approach and AO approach, we may also draw an AO-based fence by multiplying the AO bag by three and declare as outliers observations outside this fence. Figure \ref{trivarbagfence} shows the residuals with the AO bag drawn in yellow and AO fence in light blue. In Figure \ref{trivarbagloopfence} we have drawn the AO bag, loop and fence. In this case the convex polyhedrons representing the loop and fence are overlapping and in particular the loop seemingly extends further in each direction that the data is dispersed. Notably, the AO fence appears reasonably elliptical in shape, similar to its halfspace depth counterpart. \\[1ex] Under this fence based approach, 66 observations are flagged as outliers. We believe that using an AO-based fence may again fail to fully capture the shape of the data as we suspect has occurred for the halfspace depth approach. In particular, we believe this will be of the greatest concern when there is significant skewness in the data set. 

\begin{figure}[H]
	\centering 
	\begin{subfigure}{.49\textwidth}
		\centering
		\includegraphics[width=.9\textwidth]{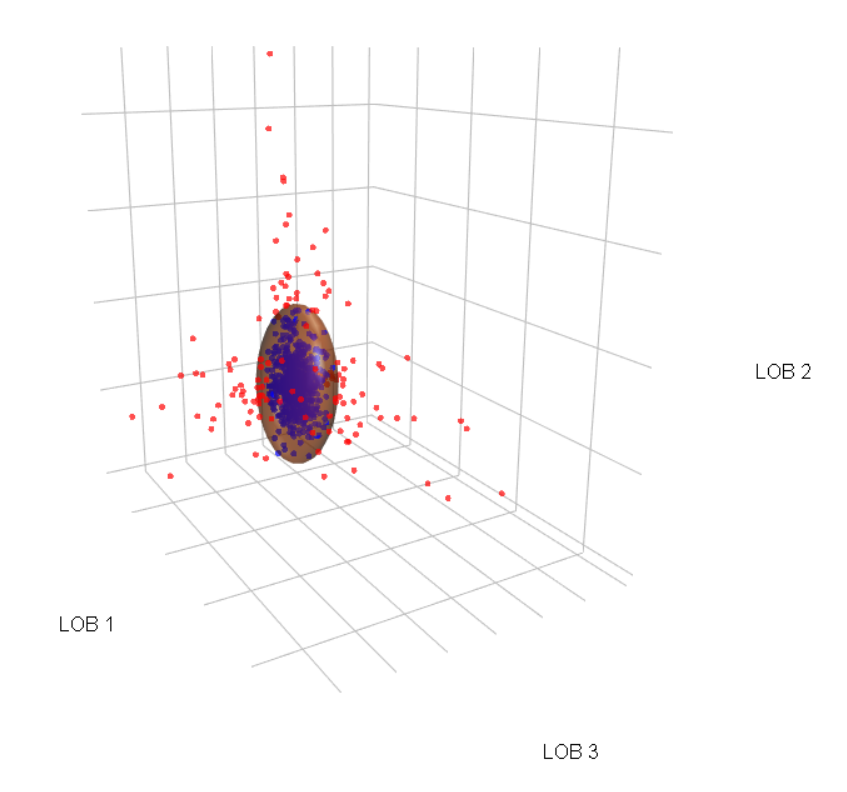}
		\caption{MCD Robust Tolerance Ellipse}
		\label{mcdellipserob}
	\end{subfigure}
	\begin{subfigure}{.49\textwidth}
		\centering	\includegraphics[width=.9\textwidth]{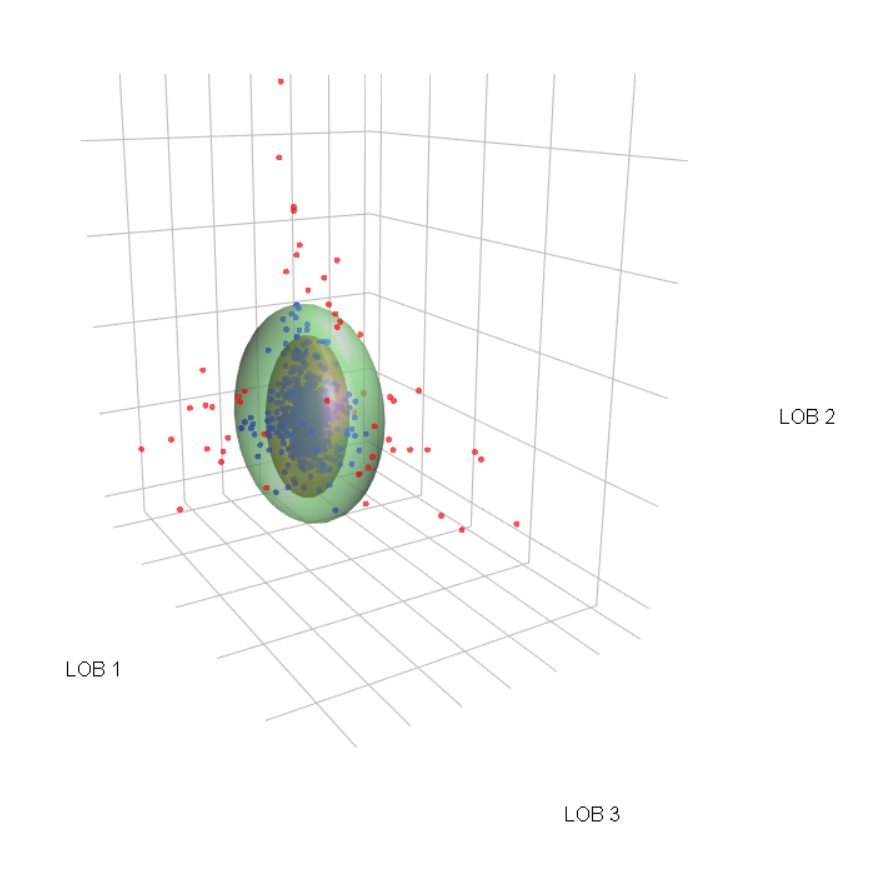}
		\caption{MCD Robust and Non-Robust Tolerance Ellipses}
		\label{mcdellipse}
	\end{subfigure}
	\caption{MCD Mahalanobis Distance detection}
	\label{trivarmcd}
\end{figure}

	\noindent Now, the MCD Mahalanobis \rev{D}istance technique which does not consider skewness in the data set at all. \rev{Analogous to tolerance ellipses as described in Section \ref{MCD}, tolerance ellipsoids have a squared distance to the central estimate of the data equal to a quantile of the $\chi^2_3$ distribution. This concept is extendable to higher dimensions where the squared distance is equal to a quantile of the $\chi^2_p$ distribution where $p$ represents the dimension of the data.}  Figure \ref{mcdellipserob} shows the residuals with the 97.5\% robust tolerance ellipsoid drawn in orange and Figure \ref{mcdellipse} shows the non-robust tolerance ellipsoid drawn in green with its robust counterpart encapsulated within it. 
\noindent This is analogous to what we saw in the bivariate case (see Figure \ref{ShBaMe12AOBagplots}). Under the MCD Mahalanobis distance approach, 120 observations were flagged as outliers. This approach saw a substantially greater number of observations detected as outliers than the other available techniques as it does not consider skewness, likely leading to misclassifications e.g. \rev{swamping} as the tolerance ellipsoid does not  extend towards the skewed data.

\subsubsection{Treatment of Outliers}
Now we turn to the treatment of outliers. Under the AO, halfspace depth and MCD Mahalanobis distance approaches, outlying observations are brought back to a relevant convex polyhedron (i.e. loop or fence) or tolerance ellipsoid.  
Under the \textit{bagdistance} methodology an alternative approach may be used which weights the degree of adjustment according to the level of outlyingness of each observation. In particular we have again considered the weighting functions given by equations \eqref{bdadj1} and \eqref{bdadj2} and set $f=\sqrt{\chi^2_3}$ and $u=10$. \rev{As noted in Section \ref{bdadjsec}, $u$ is an arbitrary selection that will be dependent on the data set under review.} Adjusted residuals under each methodology are backtransformed to give robust incremental observations. 

\subsubsection{Final Reserves and Discussion}
The time series multivariate chain-ladder technique \citep*{MeWu08} is performed on these robust observations. The results for final reserves and estimated \rev{RMSE} are given in Table \ref{trivarresults}. Note that we have scaled the presented results within each LOB \rev{to} ensure commercial data confidentiality.
	\label{outliersegments}
\begin{table}[htb]
	\begin{center}
		\resizebox{\linewidth}{!}{%
			\begin{tabular}{|c|c|c|c|c|c|c|c|c|c|c|c|} \hline
			     \multirow{2}{*}{} & \multirow{2}{70pt}{\centering \rev{No. of Outliers Detected}} &\multicolumn{2}{c|}{LOB 1} & \multicolumn{2}{c|}{LOB 2}& \multicolumn{2}{c|}{LOB 3} & \multicolumn{2}{c|}{Total} & \multicolumn{2}{c|}{\rev{Difference (\%)}} \\ \cline{3-12}
				& &Reserve&\rev{RMSE}&Reserve&\rev{RMSE}&Reserve&\rev{RMSE}&Reserve&\rev{RMSE}&\rev{Reserve}&\rev{RMSE} \\ \hline
				Original & \rev{\textbf{--}} & 900.00 & 66.62 & 1 300.00 & 112.34 & 800.00 & 74.48 & \textbf{3 000.00} & \textbf{176.69} & \rev{\textbf{--}} & \rev{\textbf{--}} \\    
				MCD & \rev{120} & 826.15 & 42.19 & 1 241.05 & 97.84 & 750.55 & 39.81 & \textbf{2 802.70}& \textbf{138.34} & \textbf{\rev{-6.58}} & \textbf{\rev{-21.71}} \\  
                HD Fence & \rev{48} & \rev{904.73} & \rev{56.61} & \rev{1 134.21} & \rev{96.41} & \rev{770.12} & \rev{47.26} & \textbf{\rev{2 843.71}} & \textbf{\rev{156.06}} & \textbf{\rev{-5.21}} & \textbf{\rev{-11.68}} \\  
                HD Loop & \rev{48} & \rev{905.34} & \rev{56.25} & \rev{1 131.76} & \rev{95.82} & \rev{765.90} & \rev{46.17} & \textbf{\rev{2 841.63}} & \textbf{\rev{155.07}} & \textbf{\rev{-5.28}} & \textbf{\rev{-12.24}} \\   
                AO Fence & \rev{66} & 903.41 & 55.39 & 1 128.94 & 95.20 & 759.14 & 41.51 & \textbf{2 834.21} & \textbf{153.22} & \rev{\textbf{-5.53}} & \rev{\textbf{-13.28}} \\  
                AO Loop & \rev{34} & 924.92 & 61.04 & 1 130.74 & 96.31 & 765.70 & 45.93 & \textbf{2 875.41} & \textbf{162.25} & \rev{\textbf{-4.15}} & \rev{\textbf{-8.17}} \\  
				AO-mrfDepth & \rev{48} & 876.42 & 57.85 & 1 256.43 & 104.58 & 764.17 & 42.68 & \textbf{2 909.42} & \textbf{160.23} & \rev{\textbf{-3.02}} & \rev{\textbf{-9.32}} \\  
				\textit{bd} (no limit) & \rev{48} & \rev{887.55} & \rev{61.79} & \rev{1 286.06} & \rev{108.27} & \rev{783.01} & \rev{54.57} & \textbf{\rev{2 961.24}} & \textbf{\rev{167.60}} & \textbf{\rev{-1.29}} & \textbf{\rev{-5.15}} \\ 
				\textit{bd} (limit) & \rev{48} & \rev{888.25} & \rev{61.82} & \rev{1 286.33} & \rev{108.29} & \rev{773.03} & \rev{54.26} & \textbf{\rev{2 960.83}} & \textbf{\rev{167.75}} & \textbf{\rev{-1.31}} & \textbf{\rev{-5.06}} \\ \hline
			\end{tabular}}
			\caption{Trivariate Reserves (Scaled values with different radix)}
			\label{trivarresults}
		\end{center}
	\end{table}
	
\noindent Under each methodology reserves have been reduced. The greatest reduction was seen when outliers were detected and adjusted using the MCD Mahalanobis distance approach. This is likely because there was a much greater proportion of observations detected as outliers under this approach. However as this methodology does not consider the skewness in the data set it may be inappropriate in this situation (likely excessive). Rather the AO technique based on the traditional cut-off value is likely more equipped to capture the shape of the data in this regard and we believe it to be the most suitable approach for this data set. The smallest reduction in reserves is given for the \textit{bd} approach where there was no upper limit on the weighting function. This is likely because under this approach outliers have been adjusted to a smaller degree than in comparison to the alternative mechanisms. Additionally, we have seen the \rev{RMSE} of reserves decrease in each case. The MCD Mahalanobis distance approach saw the greatest reduction in this metric of \rev{21.71}\% whereas the traditional AO approach (loop) saw a \rev{8.17}\% reduction. Notably the shift in the \rev{RMSE} for each approach was significantly greater than the corresponding change in reserves. This highlights that even for small adjustments in terms of the point estimate of reserves when using these robust techniques, we may see significant improvements in accuracy. 

 \section{R Codes} \label{S_Rcode}
All relevant R codes used for the bivariate and trivariate illustrations can be found at \url{https://github.com/agi-lab/reserving-robust}, along with a synthetic dataset (with features inspired by the AUSI data used in section \ref{Ndimframe}) to help replication of results.
	
\section{Conclusion}\label{conclusion}
We put forward two alternative robust bivariate chain-ladder techniques. The first technique is based on Adjusted Outlyingness and explicitly incorporates skewness into the analysis whilst providing a unique measure of outlyingness for each observation. The second technique is based on \textit{bagdistance} which is derived from the bagplot however is able to provide a unique measure of outlyingness and a means to adjust outlying observations based on this measure. Using a real bivariate data set, we illustrated how those techniques compared to existing ones. While the (inevitable) reduction in central estimates is (desirably) modest, the reduction in \rev{RMSE} is much more significant. This highlights that even for small adjustments in terms of the point estimate of reserves when using these robust techniques, we may see significant improvements in accuracy. We showed that our methodology offers material and significant improvements, compared to the techniques introduced in \citet{VeVa11}.

We then extended our framework to N dimensions and implemented all four outlier detection and treatment techniques on a real trivariate data set. The mathematical generalisation of the bivariate methodologies were not trivial. The illustration demonstrated good performance again, and results were qualitatively similar to those discussed in the bivariate example.

Through our expositions and extension to higher dimensions, we have added to the toolbox of techniques available to detect and treat outliers in multivariate reserving. We believe that the new techniques applied in this paper address some of the shortcomings of the previous approaches and should be explored as common practice when implementing robust multivariate reserving techniques in practice. 

While we focused on the multivariate time series chain-ladder model of \citet{MeWu08} for our illustrations, it should be noted that the detection and treatment methodologies developed in this paper can be applied to other reserving techniques, including in most recent machine learning actuarial procedures \citep[see, e.g.,][]{Ric18}. \rev{Notably, our work is concerned with the detection and treatment of outliers rather than finding the best reserving model for the data and we have used the chain ladder model for illustrative purposes. In the case that one proceeds to model the robustified data, back-testing of results compared to those from unadjusted data may provide useful insights into the effectiveness of the techniques.}

\section*{Acknowledgements}
\rev{Authors are very grateful for comments from two anonymous referees, which led to significant improvements of the paper. }
Earlier versions of this paper were presented at the Australian Actuaries Institute \emph{General Insurance Seminar} conference in Melbourne (Australia), as well as at an Australian  Actuaries Institute \emph{Insights} session in Sydney (Australia). The authors are grateful for constructive comments received from colleagues who attended this event and from those who read the earlier version of the paper published online to support the presentation. The authors are also grateful to Sharanjit Paddam and Will Turvey for comments on earlier drafts of the paper.
Additionally, the authors are grateful for the research assistance of Wilson Cheng, Wanzhang (Simon) Jing, and Yun Wai (William) Ho, who contributed materially to the review of the code related to the examples presented.  

This research was supported under Australian Research Council's Linkage (LP130100723, with funding partners Allianz Australia Insurance Ltd, Insurance Australia Group Ltd, and Suncorp Metway Ltd) and
Discovery Project funding scheme (DP200101859).  Furthermore, Lavender acknowledges support from the UNSW Business School Honours Scholarship. The views expressed herein are those of the authors and are not necessarily those of the supporting organisations.

\section*{References}

\bibliographystyle{elsarticle-harv}
\bibliography{libraries}

\appendix

\section*{Appendix}
\section{Data}\label{data}
	\subsection{Bivariate}\label{bidata}
	The data used to illustrate the robust bivariate chain-ladder techniques is provided in this section. It is of incremental claims for Personal (Table \ref{ShBaMe12A1}) and Commercial (Table \ref{ShBaMe12A2}) auto insurance from a major U.S. property-casualty insurer from 1988-1997. The data is taken from \citet*{ShBaMe12}.
		\begin{table}[H]
			\begin{center}
				\resizebox{0.85\linewidth}{!}{%
					\begin{tabular}{|c| c c c c c c c c c c|} \hline
						\textbf{$i$/$j$} & \textbf{1} & \textbf{2} & \textbf{3} &  \textbf{4} & \textbf{5} &  \textbf{6} & \textbf{7} & \textbf{8} & \textbf{9} & \textbf{10} \\ \hline
						\textbf{1} & 1 376 384&	1 211 168&	535 883	&313 790&	168 142&	79 972&	39 235&	15 030&	10 865&	4 086	\\ 
						\textbf{2} &1 576 278	&1 437 150&	652 445&	342 694&	188 799&	76 956&	35 042&	17 089	&12 507 & \\ 
						\textbf{3}&  1 763 277&	1 540 231&	678 959&	364 199&	177 108&	78 169&	47 391&	25 288& &\\ 
						\textbf{4} &  1 779 698	&1 498 531&	661 401	&321 434&	162 578&	84 581&	53 449& & &  \\ 
						\textbf{5} &1 843 224	&1 573 604&	613 095&	299 473&	176 842&	106 296 & & & &  \\  
						\textbf{6} &  1 962 385&	1 520 298&	581 932&	347 434&	238 375& & & & &\\ 
						\textbf{7}&  2 033 371	&1 430 541&	633 500&	432 257&  & & & & &\\ 
						\textbf{8} & 2 072 061&	1 458 541&	727 098 &  &  & & & & &\\ 
						\textbf{9} &  2 210 754	&1 517 501  &  &  &  & & & & & \\ 
						\textbf{10} &2 206 886  &  &  &  &  & & & & &\\ \hline
					\end{tabular}}
				\end{center}
				\caption{Bivariate Data Set (a) \citep*{ShBaMe12}}
				\label{ShBaMe12A1} 
			\end{table}
			\begin{table}[H]
				\begin{center}
					\resizebox{0.85\linewidth}{!}{%
						\begin{tabular}{|c| c c c c c c c c c c|} \hline
							\textbf{$i$/$j$} & \textbf{1} & \textbf{2} & \textbf{3} &  \textbf{4} & \textbf{5} &  \textbf{6} & \textbf{7} & \textbf{8} & \textbf{9} & \textbf{10} \\ \hline
							\textbf{1} & 33 810&	45 318&	46 549&	35 206&	23 360&	12 502&	6 602&	3 373&	2 373&	778	\\ 
							\textbf{2} &  37 663&	51 771&	40 998&	29 496	&12 669&	11 204&	5 785&	4 220&	1 910&\\ 
							\textbf{3}& 40 630& 	56 318	& 56 182	& 32 473& 	15 828& 	8 409	& 7 120& 	1 125 & &\\ 
							\textbf{4} &  40 475&	49 697&	39 313	&24 044&	13 156&	12 595&	2 908& & &  \\ 
							\textbf{5} &  37 127&	50 983&	34 154&	25 455&	19 421&	5 728& & & & \\  
							\textbf{6}&  41 125& 	53 302& 	40 289&	39 912& 	6 650& & & & &\\ 
							\textbf{7} &  57 515&	67 881&	86 734&	18 109 &  & & & & &\\ 
							\textbf{8} &  61 553&	132 208&	20 923 &  &  & & & & &\\ 
							\textbf{9} &  112 103&	33 250&  &  &  & & & & &\\ 
							\textbf{10} & 37 554 &  &  &  &  & & & & &  \\ \hline
						\end{tabular}}
					\end{center}
					\caption{Bivariate Data Set (b) \citep*{ShBaMe12}} 
					\label{ShBaMe12A2} 
				\end{table}  
					\subsection{Trivariate} \label{tridata}
					We use data from three different lines of business of two Australian insurers to illustrate our robust multivariate chain-ladder techniques in Section \ref{trivareg}. In particular, we investigate CTP and home insurance lines from one insurer and a CTP line from another. The data is quarterly and runs from the beginning of 2004 through 2013.
					
					The complete data set that these three lines of business have been taken from was developed as part of a Linkage Project grant awarded by the Australian Research Council (ARC) until 2016 for a project titled \textit{Modelling claim dependencies for the general insurance industry with economic capital in view: an innovative approach with stochastic processes.} It is referred to as the AUSI dataset, an acronym of the names of the project partners (Allianz Australia Insurance Ltd, UNSW Australia, Suncorp Metway Ltd, and Insurance Australia Group Ltd).
					

\end{document}